\documentclass[preprint,12pt,authoryear,review]{elsarticle}
\usepackage{graphicx}
\usepackage{tabularx}
\usepackage{amsbsy,amsmath,amsthm,latexsym,amssymb}
\usepackage{color,ulem}
\usepackage{verbatim} 
\usepackage{lineno,hyperref}
\usepackage[T1]{fontenc}
\modulolinenumbers[5]
\newcommand{\bc}{\begin{center}}
\newcommand{\ec}{\end{center}}
\newcommand{\bfr}{\begin{flushright}}
\newcommand{\efr}{\end{flushright}}

\newcommand{\be}{\begin{enumerate}}
\newcommand{\ee}{\end{enumerate}}
\newcommand{\bi}{\begin{itemize}}
\newcommand{\ei}{\end{itemize}}
\newcommand{\bd}{\begin{description}}
\newcommand{\ed}{\end{description}}

\newcommand{\eeq}{\end{equation}}
\newcommand{\bea}{\begin{eqnarray}}
\newcommand{\eea}{\end{eqnarray}}

\newcommand{\bfi}{\begin{figure}}
\newcommand{\efi}{\end{figure}}
\newcommand{\bay}{\begin{array}{l}}
\newcommand{\eay}{\end{array}}

\newcommand{\Cel}{$^\circ$C}  
\newcommand{\cref}[1]{(\ref{#1})}   

\newcommand{\ie}{\textit{i.e.}~}
\newcommand{\ca}{\textit{ca.}~}
\newcommand{\cf}{\textit{cf.}~}
\newcommand{\eg}{\textit{e.g.}~}
\newcommand{\via}{\textit{via}~}
\newcommand{\viz}{\textit{viz.}~}
\newcommand{\vs}{\textit{vs.}~}

\newcommand{\figname}{Fig.~}

\newcommand{\fignames}{Figs.~}
\newcommand{\tabname}{Table~}
\newcommand{\secname}{Section~}

\journal{}

\begin{document}

\begin{frontmatter}

\title{Nanomechanics and Pore Structure of Sodium and Potassium Geopolymer Gels: Experiments, Molecular Dynamics and Coarse-Grained Simulations}

\author[a]{Enrico Masoero\corref{cor1}}
\author[b]{Eduardo Duque-Redondo}
\author[c]{Haklae Lee}
\author[c]{Yunzhi Xu}
\author[d]{Francesca Lolli}
\author[e]{Elie Kamseu\corref{cor3}}
\author[f]{Ange-Therese Akono\corref{cor2}}

\cortext[cor1]{Corresponding author: enrico.masoero@polimi.it}
\cortext[cor2]{Corresponding author: aakono@ncsu.edu}
\cortext[cor3]{Corresponding author: kamseuelie2001@yahoo.fr}

\affiliation[a]{
    organization={Department of Civil and Environmental Engineering, Politecnico di Milano},        
    addressline={Piazza Leonardo da Vinci 32}, 
    city={Milan},
    postcode={20133}, 
    state={MI},
    country={Italy}
    }

\affiliation[b]{
    organization={Department of Physical Chemistry, University of the Basque Country (UPV/EHU)},        
    addressline={Barrio Sarriena s/n}, 
    city={Leioa},
    postcode={48940}, 
    state={Biscay},
    country={Spain}
    }

\affiliation[c]{
    organization={Department of Civil and Environmental Engineering, Northwestern University},        
    addressline={2145 Sheridan Road}, 
    city={Evanston},
    state={IL},
    postcode={60208}, 
    country={U.S.A.}
    }
  
\affiliation[d]{
    organization={School of Civil and Environmental Engineering, Georgia Institute of Technology},        
    addressline={790 Atlantic Dr NW}, 
    city={Atlanta},
    postcode={30332}, 
    state={GA},
    country={U.S.A.}
    }
    
\affiliation[e]{
    organization={Laboratory of Materials, Local Materials Promotion Authority/MIPROMALO},        
    city={Yaounde},
    postcode={2396}, 
    country={Cameroon}
    }
    
\affiliation[f]{
    organization={Department of Civil, Construction, and Environmental Engineering, North Carolina State University},
    addressline={915 Partners Way}, 
    city={Raleigh},
    postcode={27606}, 
    state={NC},
    country={U.S.A.}
    }

\begin{abstract}
The link between composition, microstructure and mechanics of NASH and KASH gels is elusive, even in pure metakaolin-based geopolymers. This article exploits molecular mechanics, coarse-grained nanomechanics and micromechanics, to interpret new experimental results from microscopy, porosimetry and nanoindentation. KASH displays a finer nanogranular structure than NASH (3 \vs 30 nm particle diameters, 5 \vs 50 nm average pore diameters), higher skeletal density (2.3 \vs 2.02 g/cm$^3$), nanoindentation moduli (9.21 \vs 7.5 GPa) and hardness (0.56 \vs 0.37 GPa) despite a higher total porosity (0.48--0.53 \vs 0.38). This suggests a stiffer and stronger solid skeleton for KASH, confirmed through predictive molecular dynamics simulations on recent and new models of NASH and KASH. The atomistic simulations inform mechanical interactions for new, coarse-grained, particle-based models of NASH and KASH. The resulting simulations predict the nanoindentation result that KASH is stiffer than equally porous NASH and the impact of formation eigenstresses on elastic moduli. 
\end{abstract}


\begin{highlights}
\item Microstructure and mechanics compared for potassium and sodium geopolymers
    \item Significant differences in particle size and pore structure from imaging and porosimetry
    \item A new atomistic model for potassium geopolymer
    \item First coarse-grained simulations of sodium and potassium geopolymers
    \item Simulations match experimental results on indentation modulus vs. porosity
\end{highlights}


\end{frontmatter}

\section{Introduction}\label{secIntro}

The mechanical properties of concrete are the complex result of heterogeneities at multiple length scales, from the nanometres of chemical components up to the metres of engineering applications. A major contribution to the mechanical properties comes from the binding phase, \ie the cement paste. Most cement produced today is Ordinary Portland Cement (OPC), but there is a drive towards new and less CO$_2$-intensive binders. \citep{habert2020environmental,miller2021achieving}. Geopolymers are one such class of inorganic polymer cement, offering an alternative to OPC. Geopolymers are obtained from raw materials rich in silicon (Si) and aluminum (Al), such as calcined clays. After dissolution in an alkaline activating solution, the solvated ions undergo polycondensation, forming a partially disordered, highly interconnected, three-dimensional network of Si and Al atoms tetracoordinated with oxygen (O) \footnote{The cations from the solution stay physisorbed in the solid structure, providing positive charges that balance the residual negative charge of tetracoordinated aluminum} \citep{davidovits2008geopolymer}. The physical and mechanical properties of the resulting geopolymers are often comparable to, and sometimes superior to, those of hardened OPC pastes. However, these properties are sensitive to the chemical composition of the raw materials, the activating solutions, and the synthesis protocol. The resulting uncertainties around performance are among the factors currently hindering more widespread use of these binders.

Typical model systems for geopolymers employ high-purity metakaolin as a raw material, activated through aqueous solutions that are based on sodium (Na), potassium (K), or a combination of both (\eg in the forms of Na$_2$SiO$_3$, NaOH, K$_2$SiO$_3$, KOH). Already these simple systems, however, involve many design parameters, such as the relative concentrations of alkali species in solution, the relative amounts of alkali in hydroxide rather than in silicate forms, the water-binder mass ratio, the Si:Al molar ratio, and the curing temperature, pressure, and humidity \citep{ribeiro2017review}. Results in the literature, therefore, are often difficult to reconcile due to differences in composition or synthesis protocols. \cite{kriven2004effect} compared the microstructure of geopolymers made from activating solutions of pure NaOH, KOH, or a combination of both. The geopolymers were cured at high temperatures and pressures to minimise the presence of macropores. \tabname\ref{tab_litrev} summarises the synthesis protocol and composition of the geopolymers in \cite{kriven2004effect} and of other ones discussed below in this section. 
\begin{table}[h]
\caption{Literature results on composition, synthesis protocol, and type of mechanical testing for potassium and sodium-based geopolymers obtained from metakaolin.}
\centering
\resizebox{\columnwidth}{!}{%
\begin{tabular}{c c c c c c c}
\hline
Reference & Si:Al (molar) & M$_2$O/Al$_2$O$_3$ 
 & H$_2$O/M$_2$O & Na/(Na+K)$^{***}$ (molar) & Synthesis protocol & Mechanical test\\
\hline\\[-0.5cm]
\cite{kriven2004effect} & 2 & 1.2 & 11 & 0 - 1 & \shortstack[c]{24h at 80\Cel \\ 1,100 psi} & Macro tests\\[0.4cm]
\cite{duxson2005microstructural} & 1.15 - 2.15 & 1 & 11 & 0 - 1 & \shortstack[c]{Sealed, 20h at 40\Cel \\ room pressure plus \\ 2 weeks at room temp.} & Macro tests \\[0.4cm] 
\cite{lizcano2012mechanical} & 1.25 - 2.50 & 1 & 11 - 13 & 0 and 1 & \shortstack[c]{Sealed, 80\Cel~for 24/48h \\ room pressure plus \\ 1 day in air} & \shortstack[c]{Macro tests and \\ microindentation} \\[0.4cm]
\cite{musil2014novel} & 2 & 1 & 11 & 0 and 1 & Room conditions & Macro tests\\[0.4cm]
\cite{sa2019amazonian} & 2 & 1 & 11 & 0.25 and 1 & Room conditions & Macro tests\\ [0.4cm]
\cite{belena2009nanoindentation} & 1.75 & 0.8 & 14 - 15 & 1 & \shortstack[c]{Sealed or unsealed \\ 25\Cel~for 24h \\ room pressure} & Nanoindentation\\[0.4cm]
\cite{nvemevcek2011nanoindentation}$^*$ & 1.22 & 1.13 & 9.45 & 1 & \shortstack[c]{Sealed at 80\Cel \\ for 12h, room pressure} & Nanoindentation\\[0.4cm]
\cite{si2020atomic}$^*$ & 1.7 - 1.88 & 1 - 1.2 & 11.3 - 12.3 & 1 & \shortstack[c]{Sealed, room conditions \\ for 24h or more} & Nanoindentation\\[0.4cm]
\cite{zhang2017multiscale} & 1.2 - 2.2 & 0.6 - 1.2 & (41-43 wt.\%)$^{**}$ & 1 & \shortstack[c]{Sealed, 23\Cel, 40-50\% RH \\ for 28 days} & Nanoindentation\\[0.4cm]
\cite{akono2019influence} & 2 & 1 & 11 & 0 & \shortstack[c]{21\Cel, room pressure\\ for 21 days} & Nanoindentation\\
\hline
\multicolumn{7}{l}{$^*$ Compositions not given in this form in the original articles, but calculated from other data provided.} \\[-0.1cm]
\multicolumn{7}{l}{$^{**}$ Water content not expressible as a molar ratio from the data provided in the original article.}\\[-0.1cm]
\multicolumn{7}{l}{$^{***}$ This ratio is 1 for activating solutions based only on Na, and 0 for solutions based on K only.}
\end{tabular}%
}
\label{tab_litrev}
\end{table}
Characteristic pore sizes from mercury intrusion porosimetry (MIP) ranged between 10 nm or less in potassium-based geopolymers (referred to as KGP hereafter) to about 50 nm in sodium-based ones (NaGP hereafter). The total MIP-accessible porosity was between 10 and 30\%, with most results around 25\%. The macroscopic compressive strength was 40 $\pm$ 5 MPa, similar for the NaGP and KGP mixes. \cite{duxson2005microstructural} measured an increase in compressive strength with Si:Al ratios going from 1 to 2, but the effect of the activator on strength was only statistically relevant at low Si:Al = 1.2, with NaGP reaching 20 MPa \vs 10 MPa for KGP. Sodium is known to be more effective in liberating silicate and aluminate monomers during geopolymerisation \citep{duxson2007geopolymer} but it is unclear whether this should raise an expectation of better mechanical properties. Despite the limited differences in compressive strength, \cite{duxson2005microstructural} reported Young's moduli that were systematically higher for NaGP at all Si:Al ratios: from 2 GPa \vs 1 GPa for KGP at Si:Al = 1.2, to 6 GPa \vs 4 GPa at Si:Al = 1.9. To our knowledge, these results have not been replicated yet and are also difficult to reconcile with the apparent densities of the samples, which the same article showed to depend weakly on the activator (actually, KGP had a slightly higher density). \cite{lizcano2012mechanical} confirmed a systematic higher apparent density in K-based geopolymers at all the Si:Al ratios they tested; nevertheless, they measured statistically equivalent compressive strengths for NaGP and KGP, except only a higher compressive strength of KGP at low Si:Al = 1.25, opposite to \cite{duxson2005microstructural}. \cite{musil2014novel} measured tensile splitting strength only, recording higher strength for KGP although within error bars. \cite{sa2019amazonian} recorded instead a significantly higher compressive strength for pure NaGP compared to geopolymers activated through a combination of Na and K (80 \vs 60 MPa). These macroscopic results show that there is still confusion about how Na- and K-based activators impact the microstructure and mechanical properties of geopolymers from metakaolin. 
To rationalize these results, experiments and computational models are now required to deepen our understanding of the nanostructure and nanomechanics of these materials \citep{chen2022pore}. 

Nano- and micro-mechanical testing help decouple the role of the geopolymer matrix from that of larger pores and inclusions. The matrix is a sodium-aluminum-silicate-hydrate gel (NASH) in sodium-based geopolymers, a potassium-aluminum-silicate-hydrate gel (KASH) in potassium-based geopolymers, and a combined KNASH gel when both activators are used \footnote{In rigorous chemical notation, these gels are \textit{Sodium and Potassium Poly(sialate-siloxo)} \citep{davidovits1989geopolymers}}. Using microindentation, \cite{lizcano2012mechanical} measured similar Young's modulus, hardness, and fracture toughness for KGP and NaGP with various Si:Al ratios, except for slightly higher Young's modulus and hardness for KGP at low Si:Al = 1.25. Accompanying SEM images, however, showed that the pores and unreacted metakaolin particles in their geopolymers were smaller than the imprint of the microindenter, thus the results were still addressing the mechanics of the composite geopolymer paste, and not only of the gel phases in the matrix. \cite{belena2009nanoindentation} used grid indentation on NaGP to isolate the elastic modulus of the NASH gel. They found an average value of 14 GPa for samples cured in sealed conditions, and 7 GPa in unsealed conditions (quite similar to the composite values of 4-5 GPa found by \cite{lizcano2012mechanical} for samples prepared in unsealed conditions at similar Si:Al ratios). \cite{nvemevcek2011nanoindentation} also used grid nanoindentation to measure the Young's modulus of the NASH gel in samples cured in sealed conditions. They found moduli between 5 and 30 GPa, with an average of 17.72 $\pm$ 4.43 GPa, which is consistent with the result from \cite{belena2009nanoindentation} despite a different composition and water content, see \tabname\ref{tab_litrev}. \cite{si2020atomic} instead obtained moduli of 7-8 GPa for nanoindented NASH gel cured in sealed conditions, whereas \cite {zhang2017multiscale} used statistical deconvolution of nanoindentation data to identify a fully reacted gel phase with moduli between 5.5 and 25 GPa. Further literature results on nanoindentation of NASH gel refer to geopolymers obtained from low-calcium fly ash as a precursor, instead of metakaolin; they also report moduli in the 4-25 GPa range \citep{das2015effective,lee2016mechanical,ma2017micro,luo2020applying,luo2020maximum}. Micromechanical data on the KASH gel are less abundant. \cite{akono2019influence} measured nanoindentation moduli between 6 and 18 GPa, mostly clustered at 8-10 GPa. These values were confirmed several times for the same mix with a similar preparation protocol \citep{akono2020fracture,chen2021influence,mahrous2023effect}. All in all, the literature data on NASH and KASH gels indicate elastic moduli that are quite robust across a range of compositions and synthesis protocols. The moduli are similar for NASH and KASH, despite the two gels significantly differing in composition and microstructure.
However the variability of the results in the literature, stemming from the different formulations and preparation protocols employed in different works, conceals trends in mechanical properties and their relationship with the underlying microstructure. Nanomechanical simulations can help clarify these aspects,  but their application to geopolymers gels is still in its infancy. Most of the current simulations used molecular dynamics to link mechanical properties with chemical composition of NASH (mainly its Si:Al ratio and water content) at the nanometre scale \citep{sadat2016molecular,lolli2018atomistic,sadat2018reactive,hou2020nanoscale,xu2021recent,liang2022review}. The only atomistic simulations on KASH thus far assumed a fully amorphous molecular structure and did not address mechanical properties \citep{kupwade2013multi}. Mesopores and micrometer structures have been barely simulated at all. \cite{lolli2018towards} proposed first concepts for such mesoscale simulations. Recent peridynamics simulations at the micrometer scale, by \cite{sadat2021atomic}, predicted excessively high Young moduli for NASH: 70-170 GPa at porosities between 0 and 40\% \footnote{\cite{sadat2021atomic} also discussed how a `homogenized' version of the same mesoporous model structures yielded a more realistic Young modulus of 21 GPa at 40\% porosity, but the details of the model were not given and the difficulty to reconcile it with the direct peridynamics simulations was acknowledged}. To date, no mesoscale simulation has addressed KASH and the general lack of micromechanical simulations of geopolymer gels is well known \citep{zuo2025modeling}.

This article systematically compares the microstructure and nanoindentation moduli of NASH and KASH gels, from NaGP and KGP geopolymer pastes that were prepared using the same protocol and equivalent compositions. Scanning and Transmission electron microscopy (SEM and TEM), MIP, and nitrogen vapor sorption (BET) characterize the microstructure of the gels in terms of size distributions of solid nanoparticles and multiscale porosity. Large microstructural differences are recorded, yet the gels have similar nanoindentation moduli, confirming the trends from the scientific literature. To explain the results, we propose a new two-scale nanomechanical model for both the NASH and KASH gels. On the first level, molecular mechanics simulations address defective versions of sodalite and leucite crystals, taken as representative structures for NASH and KASH, respectively. Mesopores are then introduced at the second level of modeling, with the gel microstructures modeled as agglomerations of spherical nanoparticles with characteristic diameters of 50 nm for NASH and 5 nm for KASH, as seen in the TEM experiments and MIP tests. The volume of pores in the model microstructures is informed by the MIP and BET data. The nanoparticles in the model interact mechanically \via harmonic bonds that are fully parametrised from the results of the molecular mechanics simulations. The resulting model microstructures produce relationships between indentation modulus and porosity that correctly predict and explain the experimental ones. Additional simulation results also clarify the effect of residual eigenstresses on the elastic moduli, showing how the proposed combination of simulations and experiments may help clarify the relationship between synthesis protocol and mechanical properties for these important materials.

\section{Methodology}\label{secMethod}
\subsection{Experimental materials and methods}\label{secExpMet}
NaGP and KGP are synthesized using the MetaMax\textregistered metakaolin, provided by BASF (BASF Corporation, Gurnee, IL) and calcined at 750$^\circ$C. Its typical content of Si and Al oxides \citep{longhi2020metakaolin,selkala2020surface} is 54 wt.\% SiO$_2$, 43.5 wt.\% Al$_2$O$_3$. The metakaolin has an average particle size of 1.2 $\mathtt{\mu}$m and a specific surface area of 1.3 m$^2$/g. 
Other raw ingredients include amorphous fumed silica (Wacker, Munich, Germany), potassium hydroxide pellets (Thermo Fisher Scientific, Waltham, MA, USA), and sodium hydroxide pellets (Thermo Fisher Scientific, Waltham, MA, USA). Before starting the synthesis procedure, the metakaolin is oven-dried at 105$^\circ$C for 24 hours to remove residual humidity and optimise its reactivity.

The target chemical formula for NaGP is Na$_2$O$\cdot$Al$_2$O$_3$$\cdot$4SiO$_2$$\cdot$13H$_2$O, with a H$_2$O/M$_2$O = 13 ratio to achieve a complete reaction of the metakaolin. For KGP, the same Si:Al and M$_2$O/Al$_2$O$_3$ ratios are targeted, but less water is needed to obtain similar workability, leading to K$_2$O$\cdot$Al$_2$O$_3$$\cdot$4SiO$_2$$\cdot$11H$_2$O. The starting point for the mix design and synthesis were previously published methods for sodium-based metakaolin geopolymer \citep{LowryKriven2010influence} and potassium-based metakaolin geopolymer \citep{akono2019influence,akono2020fracture,chen2021influence}. These protocols were designed to optimise the reactivity of the geopolymer gel and limit the presence of unreacted phases. The synthesis involves three steps. During the first step, the waterglass is mixed. For sodium geopolymer, the waterglass solution follows the proportion of 33.70 g of deionized water, 12.23 g of sodium hydroxide, and 18.75 g of fumed silica. The water glass is mixed by dissolving sodium hydroxide pellets in deionized water using a magnetic stirrer at high speed. Afterward, fumed silica is added in small batches until full dissolution is achieved. The waterglass is then left to age for 24 hours at room temperature. For potassium geopolymer, the water glass follows the proportion of 27.61 g of deionized water, 16.89 g of potassium hydroxide (from 19.87 g of pellets with 85\% purity), and 18.43 g of fumed silica. The water glass is mixed by dissolving potassium hydroxide pellets in deionized water using a magnetic stirrer at high speed. Because the dissolution is exothermic, an ice bath is used to minimise water evaporation. Afterwards, a 4-bladed IKA Rw20 overhead digital mixer is used to mix in the fumed silica at high speed and high shear. The resulting viscous slurry is left to age for 24 hours at room temperature. 

The second step involves mixing the waterglass with metakaolin using a planetary centrifugal mixer Thinky ARE 310. The waterglass and metakaolin are mixed in two steps: centrifugal mixing and vacuum degassing. The role of the vacuum degassing is to remove air bubbles and reduce the porosity. The mixing time and speed are 3 minutes and 240 g in centrifugal force for both steps. For sodium geopolymer, the mixing proportions are 64.67 g of waterglass with 35.34 g of metakaolin; for potassium geopolymer, 65.91 g of waterglass with 34.08 g of metakaolin.

The third step in the synthesis process consists in pouring the viscous slurry into lubricated cylindrical molds 76-mm tall and 20-mm round. The molds are wrapped in polyethylene films to limit water evaporation. The covered molds are left to cure in the oven at 50\textdegree~C for 24 hours. Afterwards, the samples are demolded and stored in a vacuum desiccator for further testing. Nine samples are cast for each type of geopolymer, potassium-based and sodium-based.

\tabname\ref{tab_mater} summarises the resulting composition of our samples and their synthesis protocol in a way similar to the results from the literature collected in the Introduction section: \cf \tabname\ref{tab_litrev}.

\begin{table}[h]
\caption{Composition and synthesis protocol of NaGP and KGP pastes in this study.}
\centering
\resizebox{\columnwidth}{!}{%
\begin{tabular}{c c c c c c}
\hline
Paste type & Si:Al (molar) & M$_2$O/Al$_2$O$_3$ 
 & H$_2$O/M$_2$O $^*$ & Na/(Na+K) (molar) & Synthesis protocol\\
\hline\\[-0.5cm]
NaGP & 2.09 & 1.01 & 13.25 & 1 & \shortstack[c]{Sealed, 24h at 50\Cel, room \\ pressure, then vacuum-dried}\\[0.4cm] 
KGP & 2.11 & 1.04 & 11.19 & 0 & \shortstack[c]{Sealed, 24h at 50\Cel, room \\pressure, then vacuum-dried}\\[0.4cm] 
\hline
\multicolumn{6}{l}{$^*$ H$_2$O includes moles of water released by the dissolution of the activator, 2MOH $\rightarrow$ M$_2$O + H$_2$O}
\end{tabular}%
}
\label{tab_mater}
\end{table}
%
%

\subsection{Scanning  and Transmission Electron Microscopy}\label{secSEMTEMmeth}

The structures of sodium and potassium geopolymers are characterised at the 2-500 $\mu$m length scales using an FEI Quanta 650 Environmental Scanning Electron Microscope (FEI, 153 Hillsboro, OR, USA). Before SEM analysis, the polished geopolymer specimens are rinsed in an inert oil-based solvent and cleaned using ultrasonic energy for 2 minutes to eliminate surface contamination. Due to the low conductivity of geopolymers, a low vacuum mode with an accelerating voltage of 15.00 kV is used for the entire analysis procedure. The specimens are visualised using a low vacuum Secondary Electron (LFD) Detector under the following operation conditions: a chamber pressure of 0.53 Torr, a working distance of 10.0 mm, a spot size of 5.0, and an aperture of 4 (on the scale of the instrument).

The sub-micrometer structures of NaGP and KGP are investigated using an Analytical Scanning Transmission Atomic Resolution Electron Microscope (JEOL JEM-2100 FasTEM). Samples with particle size $\le$500 nm are prepared by manufacturing dry powders of the geopolymers. Bulk geopolymer samples are milled with ethanol using an XRD-Mill (McCrone, Westmont, IL, USA) under a grinding speed of 1500 rpm to generate powder specimens. The powdered geopolymer specimens are then temporarily dispersed in ethanol using ultrasonic energy for 2 minutes and deposited on the TEM grid (Formvar/Carbon 200 mesh, made of copper). A single tilt holder is selected for mounting the TEM grid, and the TEM is operated at 200 kV. The investigation of the geopolymer grain sizes based on the TEM images using the ImageJ software \citep{igathinathane2008shape}. To this end, representative regions measuring 100 nm x 100 nm are selected, as they best showcase the granular structure of the materials. The selected regions are duplicated for grain size analysis, calibrating the scale of the images by drawing a straight line along the scale bars in the TEM images. A threshold in contrast between grains and the background is then established, and the `Smooth' and `Watershed' functions from the `Process' tab are applied to enhance the identification of grains. The `Smooth' function is particularly useful for improving the recognition of non-spherical grains in composite matrices. Maximum particle diameter limits of 60 nm for NaGP and 10 nm for KGP are applied to filter out overly aggregated and large matrix areas.

\subsection{Porosimetry}\label{MetPoro}

The pore size distributions of both NaGP and KGP are measured using mercury intrusion porosimetry (MIP), assuming the usual relationship between pore diameter and pressure \citep{washburn1921note}:
\begin{align}
d=\frac{-4 \gamma \cos(\theta)}{P}
\label{Eq-0}
\end{align}
where $d$ is the pore diameter, $\gamma$ = 485 mN/m is the surface tension of mercury, $\theta$ = 130\textdegree~is the contact angle between mercury and the geopolymer, and $P$ is the applied pressure. The pore size distribution is determined from the increase in mercury volume intruded at each pressure and from the total porosity accessible to mercury, \viz the total volume of mercury intruded during the test.

To conduct MIP, three cylindrical samples of bulk NaGP are taken from the same batch (and, similarly, three for bulk KGP); the cylinders have a 25.4-mm diameter and a 25.4-mm height. Each cylinder was cut into multiple cubes, 3 x 3 x 3 mm$^3$ in size, 
using the Techcut 4 precision diamond saw (Allied High Tech Products, Inc., Compton, CA, USA), with N-decane as the cutting fluid. The cubes are pre-dried in an oven at 50\textdegree C for 24 hours to remove any moisture. Then, several cubes are collected to form an overall sample of approximately 1 g from each of the three original cylinders. These 1-g samples are finally transferred to an Autopore V mercury intrusion porosimeter (Micromeritics, Norcross, GA, USA). The MIP testing protocol involves two steps: a low-pressure analysis ranging from 0 to 345 kPa and a high-pressure analysis from 345 kPa to 228 MPa. This allows the detection of pore diameters ranging from 500 $\mu$m to 5.6 nm. MIP data are collected at an equilibrium rate of 0.001 $\mu$L/(g $\cdot$ s) \cite{glad2015highly}. The MIP equipment also directly measures the bulk density of the samples, $\rho_b$.


Smaller pores, up to 70 nm in diameter, are characterised using nitrogen adsorption, which consists of exposing a sample to a stream of nitrogen gas at cryogenic temperatures. To prepare the samples, one 25-mm cylinder (same as in the previous paragraph) of NaGP and one of KGP are ground into fine powders, $<1~\mu$m in diameter, using an XRD mill (McCrone, Westmont, IL, USA) with ethanol as the milling fluid. Two batches of powder, 150 mg each, are collected from each ground cylinder and degassed in vacuum at 95\textdegree C for 24 h. Nitrogen adsorption is then performed using the Micromeritics 3Flex gas adsorption analyser (Norcross, GA, USA), measuring the amount of nitrogen adsorbed at different pressures. The data are used to determine the specific surface area of the material, through the Brunauer-Emmett-Teller (BET) theory \cite{brunauer1938adsorption}, and its pore size distribution, through the Barrett-Joyner-Halenda (BJH) method \cite{barrett1951determination}. 

The porosimetry data are used to estimate the total porosity $\phi$ and the skeletal density $\rho_s$ of the geopolymer pastes. For pore sizes between 5.6 and 80 nm, both the MIP and BET analyses provide pore size distributions, which do not match exactly and sometimes differ significantly. Therefore, we consider a threshold pore size $D^*$, and we read the cumulative pore volume with $D<D^*$ from BET ($v_{bet}$, per gram of material), and the cumulative pore volume with $D\ge D^*$ from MIP ($v_{mip}$, per gram of material). The total volume of pores per gram of  material is thus:
\begin{equation}
    v_p(D^*) = v_{bet}(D^*) + v_{mip}(D^*)
\end{equation}
The skeletal density is obtained from the bulk density and from the total pore volume:
\begin{equation}
    \rho_s(D^*) = \frac{\rho_b}{1-\rho_b v_p(D^*)}
\end{equation}
The total porosity is thus:
\begin{equation}
    \phi(D^*) = \frac{v_p(D^*)}{v_p(D^*)+v_s(D^*)}=\frac{v_p(D^*)}{v_p(D^*)+\rho_s^{-1}(D^*)}
\end{equation}
where $v_s(D^*)$ is the volume of the solid phase (without any porosity) per unit mass, also a function of the threshold diameter $D^*$. Varying $D^*$ between 5.6 and 70 nm, we obtain a range of values for $\rho_s$ and $\phi$ for each geopolymer sample, consisting of one MIP plus one BET result. Since we have 3 MIP and 2 BET samples for each geopolymer type (NaGP and KGP), we consider all their combinations, obtaining 6 curves of $\rho_s(D^*)$ and $\phi(D^*)$ for NaGP, and 6 for KGP.

\subsection{Nanoindentation}\label{secIndent}

Nanoindentation tests are performed on sodium geopolymer and potassium geopolymer samples using an Anton Paar NanoHardness Tester equipped with a Berkovitch diamond indenter. Before testing, the samples are polished as previously explained in \cite{akono2019influence,akono2020fracture,chen2021influence}. The indentation tests are performed in an 11$\times$11 grid with an inter-indent spacing of 25 $\mathtt{\mu}$m spanning an area of 250 $\mathtt{\mu}$m $\times$ 250 $\mathtt{\mu}$m. During each indentation, a trapezoidal vertical force history is prescribed, with a maximum load of 2 mN, a loading/unloading rate of 4 mN/min, and a holding phase of 10 s. For sodium geopolymer, the maximum penetration depth $h_{max}$ ranges from 151 nm to 971 nm, with a median value of 530 nm. As for potassium geopolymer, the maximum penetration depth  $h_{max}$ ranges from 286 nm to 723 nm, with a median value of 431 nm. Therefore, for sodium and potassium geopolymers, each indentation test probes a material volume of average radius $3h_{max}=1,500$ nm. For each indentation test, the indentation modulus $M$ and indentation hardness $H$ are calculated using the Oliver and Pharr method \citep{oliverandpharr1992}. Specifically, the equations used are:
\begin{align}
M=\frac{\sqrt{\pi}}{2}\frac{S}{\sqrt{A}};\quad H=\frac{P_{max}}{A}
\label{EQ-1}    
\end{align}
Here, $P_{max}=2$ mN is the prescribed maximum vertical force. $S$ is the slope of the $P-h$ curve upon unloading, and $A$ is the contact area at maximum depth. Before launching the test series, the contact area curve $A-h$ was calibrated using fused silica as a reference material. Therefore, each series of indentation tests yields 121 local values of the indentation modulus $M$ and the indentation hardness $H$.

The next step is to connect the indentation modulus $M$ and the indentation hardness $H$ to the microstructure of the metakaolin-based geopolymer. To this end, we use the two-level porosity multiscale thought model of geopolymer that \cite{akono2019influence} proposed. At the micrometre level, the model idealises the geopolymer matrix as a solid with a fraction $\varphi_1$ of larger pores, called `micropores' in \cite{akono2019influence} but classifiable as capillary pores according to IUPAC notation. The solid itself, however, embeds a volume fraction $\varphi_2$ of smaller pores: `nanopores' in \cite{akono2019influence}, classifiable as mesopores according to IUPAC notation. Our results from porosimetry will support this dual-porosity model that, for each pair of M and H values obtained from a nanoindenation test, allows estimating a corresponding pair of porosities $\varphi_1$ and $\varphi_2$ and therefore the corresponding, local, total porosity $\phi = \varphi_1+(1-\varphi_1)\varphi_2$. The estimate is based on the homogenisation scheme derived in \citep{akono2019influence}, which uses granular mechanics, micromechanics, and viscoplasticity theory to connect M and H to $\varphi_1$ and $\varphi_2$, as well as to three basic mechanical constants of the geopolymer skeleton: $m_s$ for the indentation modulus, $c_s$ for the cohesion and $\alpha_s$ for the friction coefficient. This is achieved through two dimensionless upscaling functions:
\begin{align}
    M=m_s\mathcal{M}\left(\varphi_1,\varphi_2, c_s,\alpha_s\right);\quad 
    H=c_s\mathcal{H}\left(\varphi_1,\varphi_2, m_s,\alpha_s\right)
\label{EQ-2}
\end{align}
that depend on the assumed underlying morphology of the material at the microscale (self-consistent homogeneously dispersed or Mori-Tanaka) and that are derived using numerical simulations as explained in \citep{akono2019influence}. To apply the model, the user is required to fix a-priori the value of the mesoporosity $\varphi_2$. In this article, we will fix $\varphi_2$ from the measured BET pore volumes per gram $v_{bet}$ and from the skeletal densities $\rho_s$ calculated from the porosimetry tests, both functions of the threshold $D^*$ discussed in the previous section:
\begin{equation}
  \varphi_2 (D^*) = \frac{v_{bet}(D^*)}{v_{bet}(D^*)+\rho_s^{-1}(D^*)}.
\label{EQ-vphi2}
\end{equation}

\subsection{Two-level simulations: atomistic and coarse-grained}\label{secSimMet}

The aim of the simulations is to predict the relationships between indentation modulus and gel porosity measured with the nanoindentation experiments. The starting point is to create realistic atomic structures for the NASH and KASH gels, at the scale of just a few nanometres. The atomic structures provide information on skeletal density and elastic moduli, which become inputs for coarse-grained simulations at the scale above. This brings the modelling to the hundredths of nanometres of nanoindentation tests, with simulations now featuring particles and pores that are akin to those measured in the microscopy and porosimetry experiments.



\subsubsection{Atomistic simulations}\label{secSimAtoMet}

For the atomic structure of the NASH and KASH gels, we create defective crystal models following the procedure in \cite{lolli2018atomistic}. All the molecular dynamics simulations in this study are conducted using the LAMMPS simulation code \citep{thompson2022lammps} and the reactive force field ReaxFF \citep{senftle2016reaxff,fogarty2010reactive,joshi2013molecular,psofogiannakis2015reaxff}. The Verlet integration method \citep{grubmuller1991generalized} with a time step of 0.1 fs is used, along with the Nosé-Hoover thermostat and barostat \citep{evans1985nose,hoover1986constant} with coupling constants of 0.1 ps and 1 ps, respectively, to control temperature and pressure during the simulations. 

\begin{figure}
    \centering
    \includegraphics[width=1\linewidth]{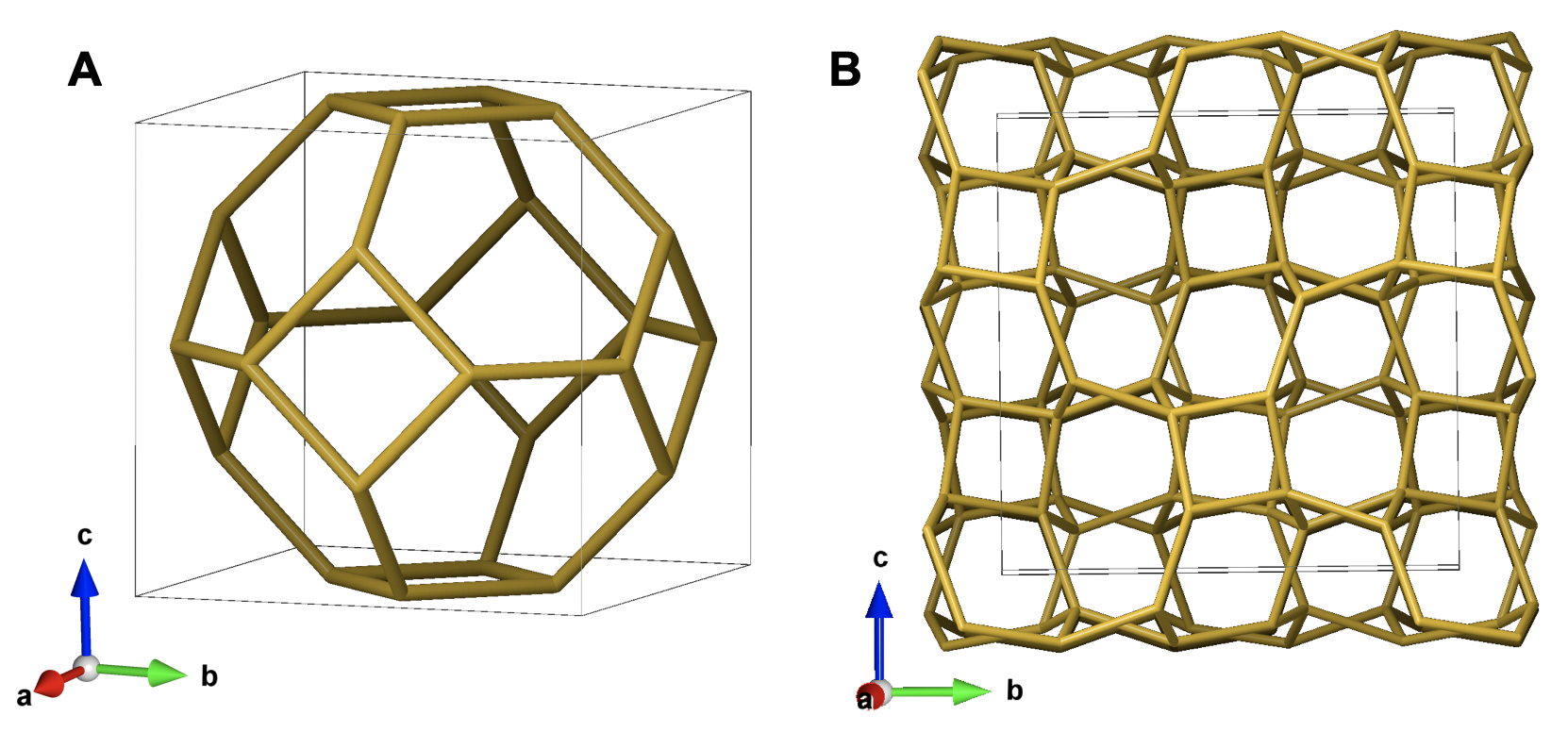}
    \caption{Baseline structures used for the construction of molecular models of NASH and KASH gels: (A) sodalite and (B) leucite frameworks.
    }
    \label{fig_baseline}
\end{figure}

For the NASH gel, the procedure to obtain a defective structure consists of partially amorphising an initially crystalline structure of sodalite zeolite (see \fignames\ref{fig_baseline}A) through the creation of vacancies, followed by the restoration of full Q$^4$ coordination of the Si atoms, the substitution of some Si with tetracoordinated Al to obtain a desired Si:Al ratio, the addition of Na$^+$ ions to restore charge neutrality, and the addition of water molecules. The resulting structure retains some trace of crystallinity despite an overall longer-range disorder, which is consistent with experimental results of X-ray diffraction, X-ray PDF, and TEM \citep{lolli2018atomistic}. For this reason, we prefer this modelling approach over adopting other amorphous model structures that have been proposed in the literature \citep{white2012molecular,sadat2016molecular,sadat2018reactive,hou2020nanoscale,xu2021recent,liang2022review}. In detail, vacancies are introduced into the sodalite framework by randomly deleting two SiO$_2$ molecules per unit cell. The resulting structures are equilibrated in the NPT ensemble (fixed pressure and temperature) at T = 300 K and P = 1 atm for 0.01 ns, followed by equilibration in the NVT ensemble (fixed volume and temperature) at T = 300 K for another 0.01 ns. After this equilibration phase, Q$^4$ connectivity is restored for all Si atoms, so that no oxygen atom will remain with dangling bonds. To achieve this, some atoms are slightly displaced by hand, and the resulting structure is equilibrated in the NPT ensemble, at T = 300 K and P = 1 atm, for 0.01 ns. This step is repeated until complete Q$^4$ polymerisation of all Si atoms is attained. The fully polymerised all-siliceous model undergoes additional equilibration in the NPT ensemble at high temperature (T = 1000 K, P = 1 atm) for 1 ns, to accelerate the relaxation of the structure before introducing subsequent modifications, \viz the partial substitution of Si by Al and the introduction of Na$^+$ and water molecules, detailed later.


For the KASH gel, the atomic structures are produced following an analogous procedure as for the NASH gel above. Also in this case the objective is to obtain disordered structures which retain traces of short-range crystallinity, as opposed to other fully amorphous arrangements proposed in the literature \citep{kupwade2013multi}. The starting point, in this case, is the crystal framework of leucite (see \fignames\ref{fig_baseline}B), because XRD experiments reveal peaks in KASH gels that are typical of this type of zeolite (especially when brought to high temperature: see \cite{bell2008x}). X-ray PDFs show no long-range peaks, indicating that crystallinity is only local and it is lost within a few nanometres; this is similar to how NASH gels are locally akin to sodalite zeolites, but disordered in the longer range. All this justifies the adoption of the same modelling protocol as for the NASH gel above, only starting from a leucite framework instead. The only difference is that four SiO$_2$ molecules per unit cell are initially removed to create vacancies, instead of two for the NASH gel above; this is because the leucite unit cell features more Si atoms than a sodalite cell.

The equilibrated defective silicate structures, obtained starting from the sodalite and leucite frameworks, are then turned into alumino-silicate structures by replacing a few Si atoms with Al atoms. Two Si:Al ratios are targeted: 1.4 and 2.0 for both NASH and KASH structures; these are the extremes of the Si:Al range for which good mechanical properties have been experimentally recorded, \eg see \cite{duxson2005microstructural}. This allows us to assess whether the Si:Al ratio has any significant impact on mechanical properties at the nanometre scale.
When replacing Si with Al, care is taken to comply with the Loewenstein principle \citep{davidovits1991geopolymers}, \ie avoid the presence of adjacent Al tetrahedra linked to each other. Tetracoordinated Al atoms introduce a negative charge therefore, to keep the systems electroneutral, Na or K cations (called M in general) are added randomly using the Packmol software \citep{martinez2009packmol}, thus setting the M/Al ratio to 1. Packmol is also used to add water molecules at random positions. Following the approach in \cite{lolli2018atomistic} and \cite{lolli2021early}, the water molecules to add to the atomic structures are identified as the strongly bound water molecules recorded during the drying experiments in \cite{kuenzel2012ambient}. For the NASH gel, this leads to H$_2$O/Na ratios of 1.55 and 2.90 respectively for Si:Al ratios of 1.4 and 2. The content of strongly bound water is significantly lower in KASH, with the results in \cite{kuenzel2012ambient} indicating H$_2$O/K ratios of 0.6 and 1.45 for Si:Al ratios of 1.4 and 2. This is due to the higher charge density of Na compared to K, which hinders the removal of water molecules from its hydration shell. The model gels structures, now featuring the desired Si:Al, M/Al, and H$_2$O/M ratios, are finally relaxed \via energy minimisation using the Polak-Ribière conjugate gradient algorithm \citep{polak1969note} and then \via MD simulations in the NPT and NVT ensembles, at P = 1 atm and T = 300K, for 5 ns in each case.

\begin{figure}
    \centering
    \includegraphics[width=1\linewidth]{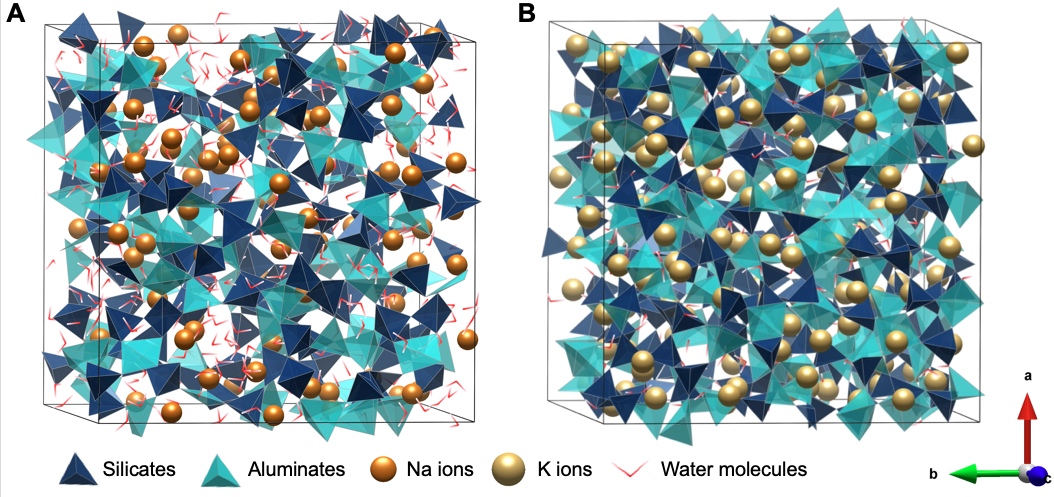}
    \caption{Atomic-scale structures of the defective models of (A) NASH and (B) KASH gels with Si:Al ratios of 1.4. 
    }
    \label{fig_AM}
\end{figure}

The resulting atomic models of NASH and KASH are shown in \fignames\ref{fig_AM}. These model structures are then characterised in terms of structure and mechanical properties. Skeletal densities are simply computed by summing the mass of all the molecules in a simulation box (SiO$_2$, AlO$_2^-$, H$_2$O, and Na$^+$ or K$^+$) divided by the volume of the box. XRD and X-ray PDFs are computed using the VESTA \citep{momma2006vesta} and TRAVIS \citep{brehm2020travis} software respectively. Mechanical properties are extracted from stress-strain curves, obtained by subjecting the model structures to uniaxial deformation. The deformation protocol consists of a series of 1\% strain steps imposed in one direction as a percent of the initial box size, whilst the perpendicular directions to the deformation are kept fixed. After each deformation step, the atomic positions are relaxed using the Polak-Ribière conjugate gradient energy minimisation algorithm \citep{polak1969note}. After relaxation, the virial normal stresses are computed (shear stresses are negligible). The initial slope of the stress-strain curve in the direction of the imposed deformation provides the indentation modulus $M$, whereas the normal stresses in the perpendicular directions are used to compute the Poisson's coefficient $\nu$ and thus the Young's modulus $E$: see \cite{lolli2018atomistic} for details. The tensile strain at rupture is also determined by analysing the stress-strain curves; it corresponds to the strain at which the stress experiences a sudden decline after reaching its peak. This critical point marks the material's breaking point, indicating the onset of significant deformation until complete fracture occurs. The full process of uniaxial deformation is repeated along the three Cartesian coordinates, thus three stress-strain curves and three sets of values of elastic constants and tensile strength are obtained for each model structure of NASH and KASH.


\subsubsection{Coarse-grained simulations}\label{secSimCGMet}

\begin{figure}
    \centering
    \includegraphics[width=1\linewidth]{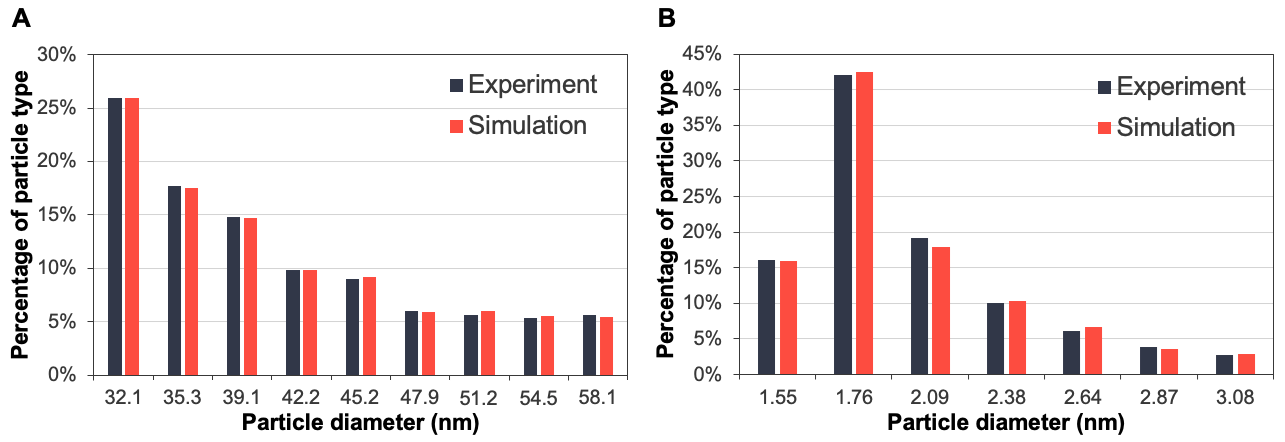}
    \caption{Comparison of experimental and simulation grain size distributions of (A) NASH and (B) KASH gels.}
    \label{fig_GSDcomparison}
\end{figure}

Mesoscale models of NASH and KASH gels are built following a similar procedure as in \cite{masoero2020nanoparticle}, where the gel structures are idealised as a cohesive agglomeration of spherical nanoparticles. The particle size distributions are informed by our SEM and TEM experiments, which indicate diameters in the 32.1-58.1 nm range for NASH and 1.55-3.08 nm for KASH. SEM and TEM images only provide information about the projected area of particles, rather than their actual diameters; to determine the diameters we assume that the particles are spherical and that the SEM and TEM sections cross them in the middle, so that the diameters are estimated from the area of a circle A = $\pi r^2$. This provides upper bound values of particle sizes, since not all particles are actually intersected through their centres. The particle size distributions obtained from the experiments are shown in \figname\ref{fig_GSDcomparison}. 

To create the mesostructures, we employ a space-filling algorithm to place particles into a cubic simulation box with a linear size of approximately 20 times the average particle diameter: \ca 1,000 nm for NASH and 50 nm for KASH. In the space-filling algorithm, nine particle sizes are initially defined within the range indicated by the TEM experiments: 58.1, 54.5, 51.2, 47.9, 45.2, 42.2, 39.1, 35.3, and 32.1 nm for NASH and seven particles sizes for KASH: 3.08, 2.87, 2.64, 2.38, 2.09, 1.76, and 1.55 nm. The algorithm first places $N_1$ particles of the largest size at random locations. If any two particles overlap by more than $(1-\xi) D$ ($D$ is the diameter of the particles and $\xi$ is a user-defined parameter that will be used later to control the intensity of eigenstresses in the system), one of the overlapping particles in the pair is deleted. The remaining particles undergo energy minimisation using the conjugate gradient (CG) approach in LAMMPS. The algorithm then proceeds with inserting $N_2$ particles of slightly smaller diameter, also at random locations in the simulation box. If a just-inserted particle overlaps by more than $(1-\xi) D$ with any other particle ($D$ is the average diameter of the overlapping pair), the smallest particle in the pair is deleted. Energy minimisation of the entire system follows again. The same steps of random insertion, removal of overlapping particles, and energy minimisation are repeated for all the other predefined particle sizes, with values of $N_i$ decided by the user of the algorithm. The whole procedure is repeated in a loop until a user-specified packing volume fraction $\eta$ is reached, with $\eta = 1 - \phi$ and $\phi$ the porosity of the model structure. By using a small $\xi$, \viz by allowing more overlapping between particles, we can achieve high packing densities but this generates large eigenstresses in the system (because of the mechanical interactions between particles discussed below). The particle size distribution, instead, can be tuned through the $N_i$ parameters. 

\begin{figure}
    \centering
    \includegraphics[width=1\linewidth]{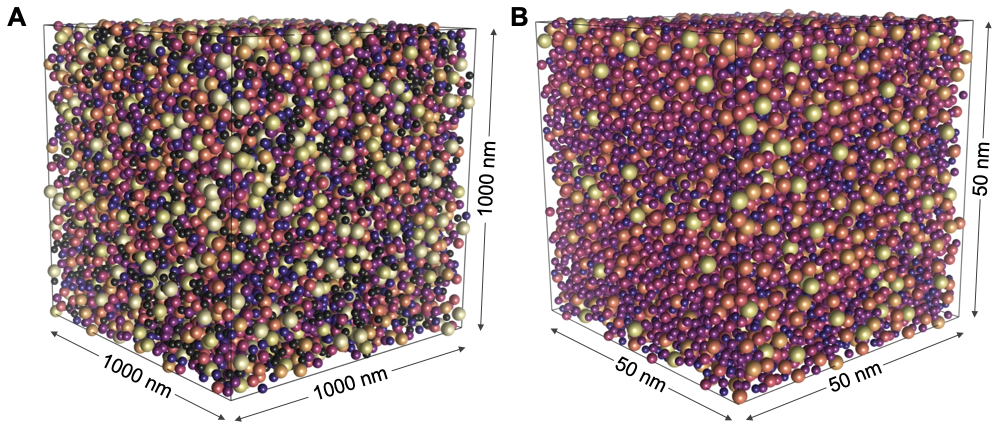}
    \caption{Mesoscale model structures of (A) NASH and (B) KASH gels with solid fractions $\eta$ of 0.67 and 0.62 respectively. The darker the colour of the spheres, the smaller their diameter.}
    \label{fig_MM}
\end{figure}

\figname\ref{fig_GSDcomparison} shows that our packing algorithm can produce particle size distribution that match well those estimated from microscopy experiments. \fignames\ref{fig_MM} shows two typical configurations of agglomerated particles produced in this work, with 10,000 to 21,000 polydisperse particles for the NASH gel, (depending on the target $\eta$), and with 12,000 to 27,500 polydisperse particles for KASH.

The mechanical interactions between particles are dictated by a pairwise, size-dependent, harmonic potential, which is rather common in nanoparticle simulations \citep{masoero2020nanoparticle,harrison2018review,mishra2017cemff}: $U_{ij} =\frac{1}{2}k(r_{ij}- D_{ij})^2-U_{0,ij}$, where \textit{r$_{ij}$} is the distance between particles \textit{i} and \textit{j}, \textit{D$_{ij}$} is their average diameter, and $k=EA_{ij}/D_{ij}$. The contact area between two particles is $A_{ij}=\frac{\pi}{4}D_{ij}^2$, with \textit{E} being the elastic modulus obtained from simulations at the atomic scale. Additionally, a cutoff is implemented, which imposes $U_{ij} = 0$ when $r_{ij}\geq r_u$, with $r_u = \varepsilon_u D_{ij}$ where $\varepsilon_u$ is the strain at tensile failure, also obtained from the atomistic simulations. In the Results section, atomistic simulations will provide $E$ and $\varepsilon_u$ values for NASH and KASH gels, both for Si:Al = 1.4 and 2. The values adopted in the mesoscale simulations are those from the model which is closest to the experimental Si:Al ratio, 2.0, \viz $E = 59.42$ GPa and $\varepsilon_u = 0.217$ for NASH, and $E = 65.97$ GPa and $\varepsilon_u = 0.285$ for KASH.

Before subjecting the model mesostructures to mechanical tests, the average axial stresses in all three directions are set to zero. To this end, the lengths of the box in the x, y, and z directions are adjusted, which leads to updated values of packing density $\eta$ for each model mesostructure generated with the aforementioned packing algorithm. The average stress components in the $a$ and $b$ directions are computed using the virial method for pairwise interaction potentials implemented in LAMMPS \citep{thompson2022lammps}:
\begin{equation}
    \sigma_{ab}= -\frac{1}{2V} \sum_i\left[\sum_{j\neq i} \left( r_{a,i}F_{ij,b,i}+r_{a,j}F_{ij,b,j} \right)\right]
\end{equation}
$V$ is the volume of the simulation box, $i$ and $j$ count all the particles in the box, $r_{a,i}$ is the $a$ coordinate of particle $i$, $F_{ij,b,i}$ is the $b$ component of the interaction force on particle $i$ coming from its interaction with particle $j$ ($F_{ij}$ is positive when attractive, negative when repulsive). A deformation protocol is then employed, similar to the one in the atomistic simulations, with uniaxial strain in one direction whilst the perpendicular directions are kept fixed. However, instead of applying a fixed strain increment, an exponentially increasing strain increment is adopted here, which allows finely resolving the initial linear regime at small strains and then speeding up the deformation process until reaching an 80\% axial strain that ensures full mechanical failure. 
The uniaxial deformation is replicated across all three Cartesian coordinates, yielding three stress-strain curves obtained for each mesoscale model of NASH and KASH with different packing fractions. The initial slope of these stress-strain curves is used to estimate the indentation modulus $M$.

\section{Results}\label{secRes}

This section first reports the results from the experimental characterisation: SEM and TEM images, MIP and BET porosimetry, and nanoindentation. Simulation results follow, which use some of the experimental results along with elastic properties from the model atomic structures of NASH and KASH gel, to predict and rationalise the results of nanoindentation on Na and K-based geopolymers.

\subsection{Electron Microscopy}\label{secResImage}

\begin{figure}[h]
    \centering
    \includegraphics[width=\linewidth]{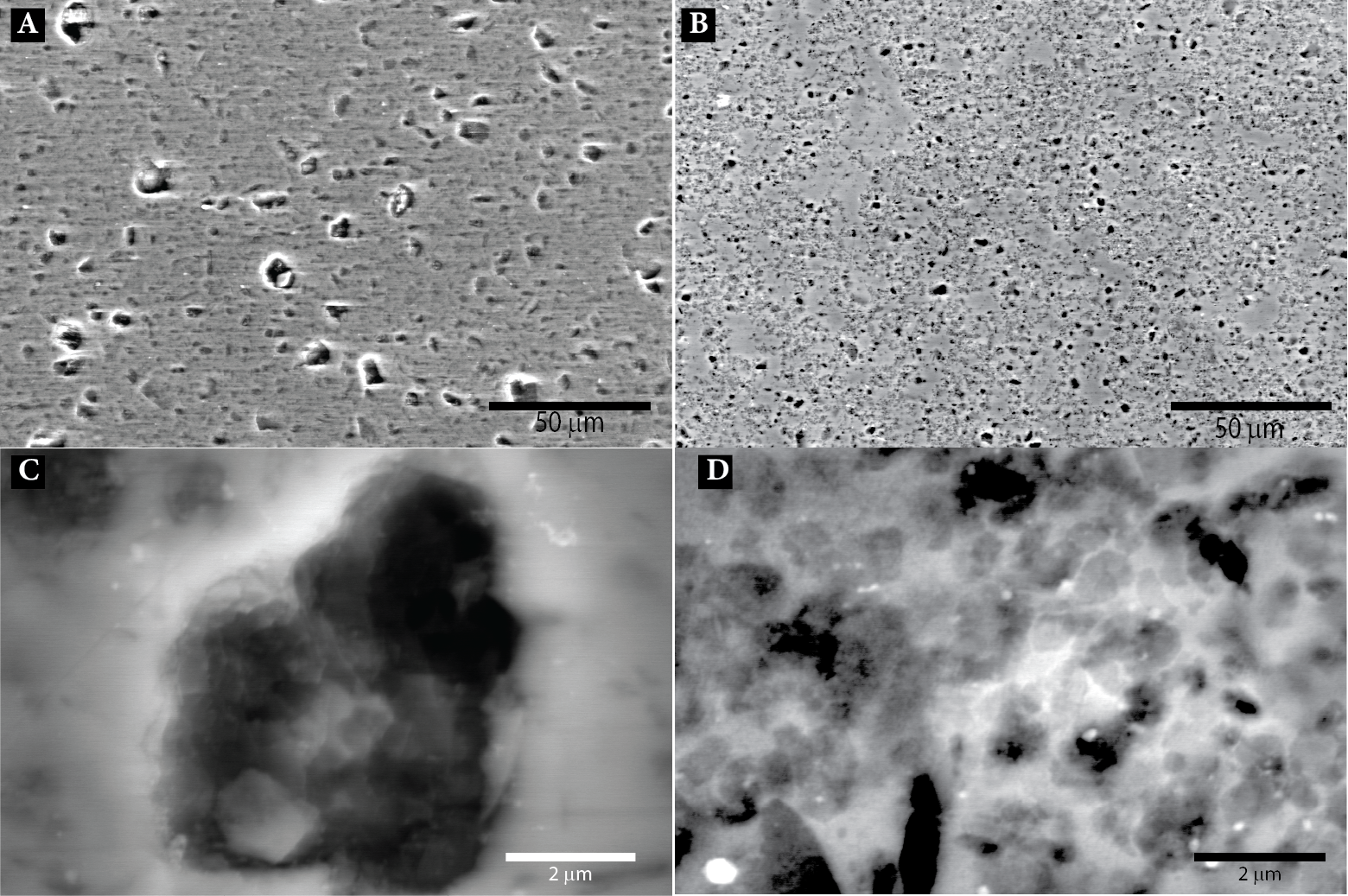}
    \caption{SEM images of (A, C) NaGP and (B, D) KGP pastes.  
    }
    \label{fig_SEM}
\end{figure}

\begin{figure}[h]
    \centering
    \includegraphics[width=0.87\linewidth]{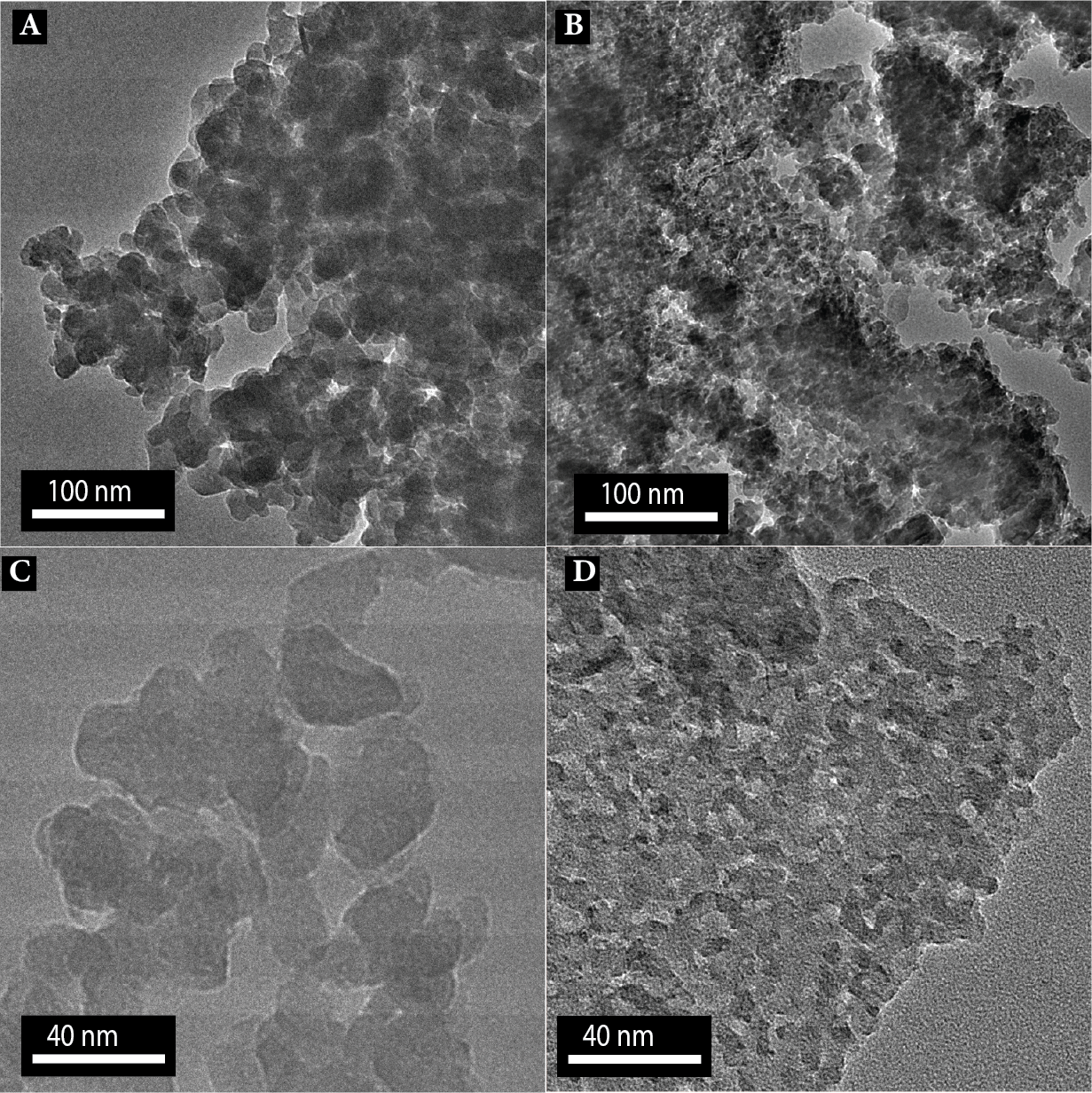}
    \caption{TEM images of (A, C) NaGP and (B, D) KGP pastes. 
    }
    \label{fig_TEM}
\end{figure}

\begin{figure}[h]
    \centering
    \includegraphics[width=\linewidth]{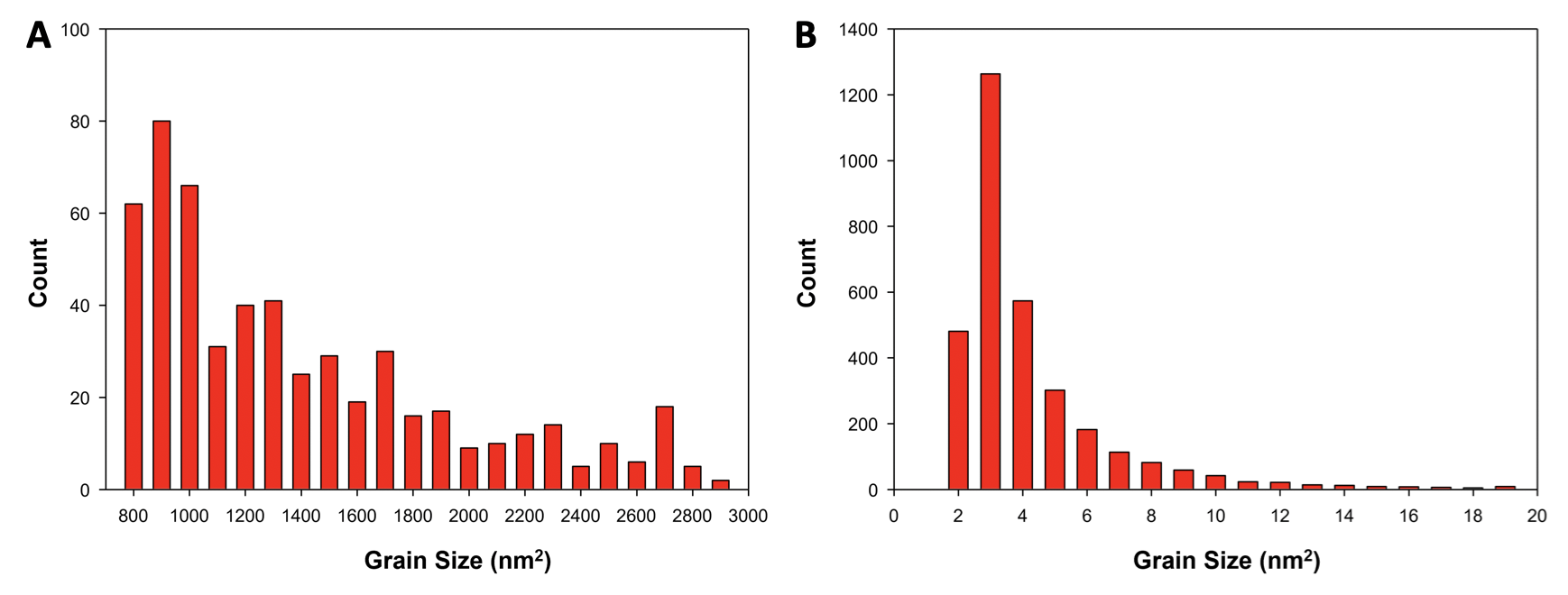}
    \caption{Areal particle size distributions of (A) NASH and (B) KASH based on the TEM image analysis results.}
    \label{fig_grain_PSD}
\end{figure}

\figname\ref{fig_SEM} shows SEM images of the structure of our NaGP and KGP pastes. 
%
%
At low magnification (500x), in \fignames\ref{fig_SEM}.A and \ref{fig_SEM}.B, the NaGP paste exhibits more and larger pores compared to the KGP samples. In particular, the macropores in the NaGP pastes have sizes of $\sim 30 \mu$m, as opposed to the less than 10 $\mu$m in the KGP pastes. Solid-void interfaces for these pores are shown at high magnification (20,000x) in  \fignames\ref{fig_SEM}.C and \ref{fig_SEM}.D. The images indicate a coarser surface structure of the NaGP paste, as opposed to the finer and somewhat fibrillar morphology of the pore surface in the KGP.
%
%
This is confirmed by TEM images, such as those in \figname\ref{fig_TEM}, which show the nanogranular structure of both geopolymers, with nanoparticle sizes of \ca 30 nm in the NASH gel and of \ca 5 nm or less in the KASH gel (\cf similar results in the literature, \eg in \cite{autef2013influence}).
%
%
%
%
The histograms in \figname\ref{fig_grain_PSD}, obtained through the ImageJ analyses detailed in \secname\ref{secSEMTEMmeth}, quantify the particle size distributions in the NASH and KASH gels. 
%
%
%
%
The distributions clearly reflect the difference in particle sizes for NASH and KASH gels, already evident from the TEM images. Indeed, the average areal grain sizes are \ca 1340 and 4 nm$^2$ respectively for NASH and KASH gels. The shapes of the distributions in \figname\ref{fig_grain_PSD} are instead somewhat similar, in that the most abundant particles are those closer to the smallest size in the range, with a long tail of larger particles. 

\subsection{Pore structure}\label{secResPoro}

\begin{figure}
    \centering
    \includegraphics[width=1\linewidth]{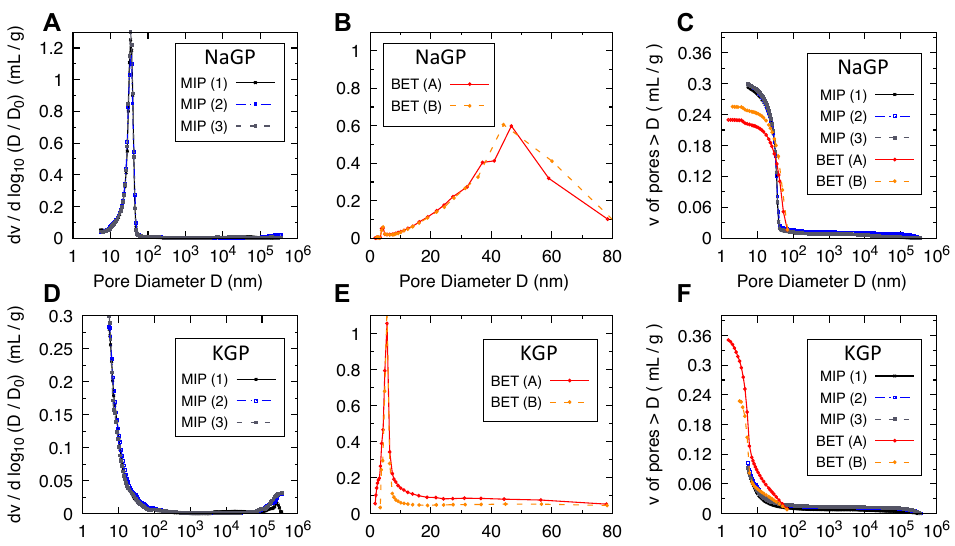}
    \caption{Pore size distributions of (A-C) NaGP and (D-F) KGP: log-differential volume distributions from (A,D) mercury intrusion porosimetry, MIP, and (B,E) nitrogen sorption tests, BET; (C,F) cumulative pore volume distributions from both MIP and BET tests. The measure total pore volumes are the leftmost points in C and F, which indicate a range of 0.23 - 0.36 mL/g for both NaGP and KGP samples, similar to the 0.22 - 0.3 mL/g range obtained by \cite{kriven2006microstructure} depending on the proportion of Na and K in the activator.
    }
    \label{fig_pores}
\end{figure}

The pore size distributions of the gels in NaGP and KGP, exhibit significant differences, as depicted in \figname\ref{fig_pores}. The peak in the distribution for NaGP lies in the 20 to 80 nm diameter range, as consistently indicated by both the MIP and BET results in \fignames\ref{fig_pores}.A and \ref{fig_pores}.B. In contrast, the log-differential pore volume distributions of KGP peak around diameters D of 5 nm; this challenges the resolution of MIP in \figname\ref{fig_pores}.D, but is well captured by nitrogen sorption (BET) in \figname\ref{fig_pores}.E. In both cases, it is thus the gel, NASH and KASH respectively, which controls the pore size distributions of the pastes. There is little variability in the characteristic pore sizes emerging from repeated tests on statistically equivalent samples. The difference in characteristic pore sizes between NaGP and KGP, \ca 50 and 5 nm respectively, is consistent with the TEM images in \figname\ref{fig_TEM}.

\fignames\ref{fig_pores}.C and \ref{fig_pores}.F show the cumulative pore volume distributions for NaGP and KGP, from both the MIP and the BET tests. For NaGP, the cumulative distributions in \figname\ref{fig_pores}.C display systematic inconsistencies between the MIP and BET results. Both cumulative distributions show a sharp increase in pore volumes at diameters $D$ between 20 and 80 nm, which is consistent with the log-differential distributions in \fignames\ref{fig_pores}.A and \ref{fig_pores}.B. However, the sharp increase occurs at higher $D$ values in the BET results than in those from MIP. Also, the MIP tests measure a higher total pore volume than the BET tests, if both are extrapolated down to $D = 1$ nm: \ca 0.24 mL/g from BET against over 0.3 mL/g from MIP. The opposite should be the case, if both the BET and MIP tests were able to sample all the available pores, since BET can reach smaller pores than MIP. Both these counterintuitive results may originate from an ink-bottle effect if a significant fraction of large mesopores in NaGP, with $D \gtrsim 50$ nm, is enclosed as pockets within a network of smaller pores, called necks. In such a scenario, the ink-bottle effect would be less effective in the BET tests because nitrogen is generally better than mercury in accessing smaller necks, and because some of the necks and pockets would be disrupted by the grinding process, as part of sample preparation for BET. As a result, BET better detects the large mesopores (hence the right shift of the BET curves in \figname\ref{fig_pores}.C) but only those that survive the grinding process (hence a lower total pore volume when $D\rightarrow 1$ nm than from the MIP tests). By contrast, MIP only detects the pockets of larger mesopores when the mercury pressure is sufficiently high to penetrate the small necks. Therefore, MIP underestimates the presence of large mesopores, misattributing their volume to smaller pores. However, since MIP is conducted on the original unperturbed bulk samples and since the NaGP has very few pores with $D < 5.6$ nm (the smallest diameter that MIP can detect), the MIP results in \figname\ref{fig_pores}.C do eventually access most of the pore volume \footnote{For KGP, instead, the BET results in \figname\ref{fig_pores}.F probably capture the entire pore volume because the grinding process, in that case, should not alter much the volume of small pores that are prevalent in KGP.}. 

\begin{figure}
    \centering
    \includegraphics[width=0.93\linewidth]{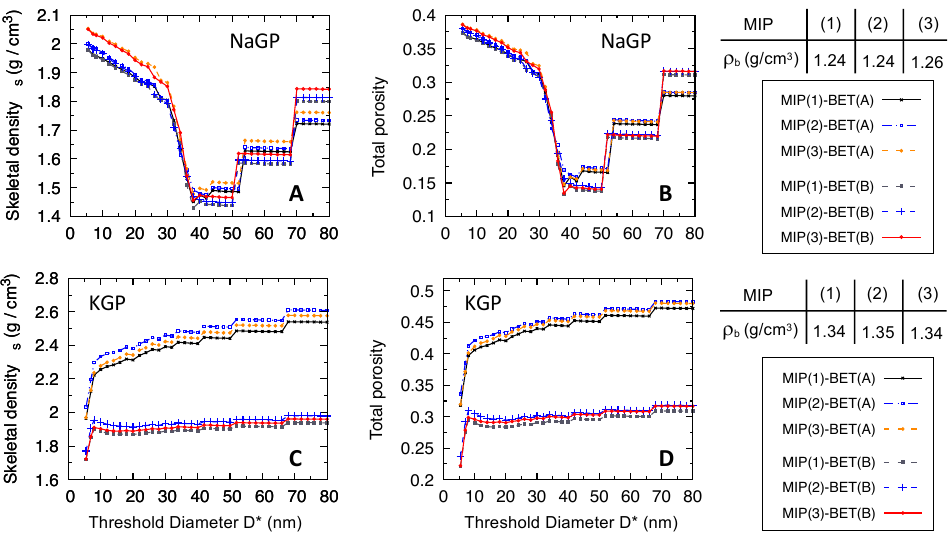}
    \caption{Skeletal density $\rho_s$ and total porosity $\phi$ of: (A,B) NaGP, and (C,D) KGP, as functions of the threshold pore diameter $D^*$ (the porosity data for pores with $D < D^*$ are taken from the BET datasets, for $D \ge D^*$ from MIP). The right side of the figure shows the legends with all the combinations of MIP and BET results in \figname\ref{fig_pores}, as well as the values of bulk densities $\rho_b$ measured during the MIP experiments on the NaGP and KGP pastes.  
    }
    \label{fig_poroskele}
\end{figure}

Based on the proposed ink-bottle effect, the best values of skeletal density $\rho_s$ and total porosity $\phi$ for the NaGP samples are obtained when taking $D^* = 5.6$ nm, which is the resolution limit of MIP (see \secname\ref{MetPoro} for the definition of the threshold pore diameter $D^*$). This means that the whole pore volume recorded by MIP in \figname\ref{fig_pores}.C is considered, whereas the BET results are only used to complement the MIP data in the 1 to 5.6 nm range. For $D^* = 5.6$ nm, \fignames\ref{fig_poroskele}.A and \ref{fig_poroskele}.B indicate $\rho_s = 1.97-2.07$ g/cm$^3$ and $\phi = 0.37-0.39$. Compared to the results for KGP in the same figure, these values are in the lower range of $\rho_s$ and mid-range for $\phi$. For larger $D^*$, the curves in \fignames\ref{fig_poroskele}.A and \ref{fig_poroskele}.B exhibit significant variations, which reflect the inconsistencies between MIP and BET stemming from the ink-bottle effect.

The same analysis of cumulative pore volume distributions, skeletal density, and total porosity is repeated for the KGP pastes. In this case, the two techniques yield consistent trends for the KGP samples, with the BET distributions extending those from MIP down to the nanometre range: see \figname\ref{fig_pores}.F . There is however variability between the two repeated BET tests, with BET(A) indicating that pores in the 10-100 nm range account for a larger pore volume than suggested by the MIP tests and by the BET(B) test. Also, BET(B) records a smaller total pore volume than BET(A): \ca 0.26 mL/g against \ca 0.36 mL/g when extrapolating the curves in \figname\ref{fig_pores}.F  down to $D = 1$ nm. To account for these differences, \fignames\ref{fig_poroskele}.C and \ref{fig_poroskele}.D compute the total porosity $\phi$ and solid skeletal density $\rho_s$ of KGP for all the possible combinations of MIP and BET results presented here. $\phi$ and $\rho_s$ are computed as functions of the threshold diameter $D^*$, below which the pore volumes are taken from BET, and above which from MIP: see \secname\ref{MetPoro} for details. For $D^* \leq 10$ nm, the $\phi$ and $\rho_s$ values of KGP change steeply: here MIP is close to its resolution limit, so the results for $D^* \geq 10$ nm are more reliable in this case. Another message from \fignames\ref{fig_poroskele}.C and \ref{fig_poroskele}.D is that the computed $\phi$ and $\rho_s$ significantly change when using the results from the BET(A) or the BET(B) tests. There is no strong reason to prefer one BET test result over the other; the only noteworthy fact is that the smallest pore size sampled in the BET(B) test, \ca 3 nm, is larger than the one sampled in the BET(A) test, \ca 1 nm (see \figname\ref{fig_pores}.F) hence the BET(B) test underestimates the total pore volume, and therefore also $\phi$ and $\rho_s$, by \ca 10\% (this means that, for $D^* = 80$ nm for example, one may consider $\rho_s \approx 2.2$ g/cm$^3$ and $\phi \approx 0.35$ for KGP instead of $\rho_s \approx 2$ g/cm$^3$ and $\phi \approx 0.32$ as shown in \fignames\ref{fig_poroskele}.C and \ref{fig_poroskele}.D). Based on this interpretation, the results in \fignames\ref{fig_poroskele}.C and \ref{fig_poroskele}.D indicate that KGP has a skeletal density $\rho_s$ between 2 and 2.6 g/cm$^3$ and a total porosity $\phi$ between 0.3 and 0.48.

All in all, the analysis of the pore structures has indicated that: (i) KASH gels have higher average skeletal density $\rho_s$ than NASH gels: 2.3 $\pm$ 0.3 g/cm$^3$ \vs 2.02 $\pm$ 0.05 g/cm$^3$; (ii) NASH and KASH gels have very different textures, with prevalent pore diameter in the order of 50 nm for NASH and 5 nm for KASH, nevertheless (iii) NASH has total porosity $\phi = 0.38 - 0.39$ whereas the porosity results for KASH are not conclusive in that a wide range of possible values has emerged, between 0.32 and 0.48; the nanoindentation results in the next section will indicate porosity values towards the upper bound of this range.

\subsection{Nanoindentation}\label{secResIndent}

The nanoindentation tests provided a set of 121 pairs of indentation modulus $M$ and hardness $H$ for both NASH and KASH gels. The micromechanical model in \secname\ref{secIndent} associate to each $M-H$ pair a best-fit value of local total porosity $\phi$, provided an input the value of mesoporosity $\varphi_2$. 

For our NaGP samples, the experimental porosimetry results in the previous section have indicated $D^* = 5$ nm as a recommended value, leading to a negligible value of $v_{bet}$ (0.007-0.008 mL/g \vs a total pore volume of 0.22-0.32 in \figname\ref{fig_pores}). Therefore, for NaGP we consider $\varphi_2 = 0$, meaning that a single distribution of capillary pores is sufficient to describe the microstructure of the NASH gel. \tabname\ref{tab_indent} shows the average values of $M$ and $H$ from nanoindentation, as well as the average values of capillary porosity $\varphi_1$ and total porosity $\phi$ from the micromechanical model. For NaGP, the average $M$ is in line with the results from literature presented in the Introduction section. The average value of $\phi = 0.373$ falls  within the narrow 0.37 -- 0.39 range from porosimetry in the previous section.
\begin{table}[h]
\caption{Microstructural and mechanical parameters obtained from the nanoindentation tests and their intepretation through the micromecanical model in \secname\ref{secIndent}.}
\centering
\resizebox{\columnwidth}{!}{%
\begin{tabular}{l | c c c | c c c c}
\hline
& \multicolumn{3}{c|}{\textbf{Micromechanical model inputs}} & \multicolumn{4}{c}{\textbf{Model and nanoindentation results$^\dag$}}\\[-0.1cm]
 & $v_{bet}(D^*)$ & $\rho_s(D^*)^{\dag\dag}$
 & $\varphi_2$ &  $\varphi_1$ &  $\phi$  & $M$ &  $H$\\[-0.1cm]
Dataset name &  (mL/g) & (g/cm$^3$) & & & & (GPa) & (GPa) \\
\hline
NaGP & 0.007 - 0.008 & 2.02 $\pm$ 0.05 & 0 & 0.373 & 0.373 &  7.5 & 0.37\\[0.0cm] 
KGP\_BET(A)\_$D^*10$nm & 0.26 & 2.29 $\pm$ 0.03 & 0.38 & 0.247 & 0.525 & 9.12 & 0.56\\[0.0cm] 
KGP\_BET(A)\_$D^*80$nm & 0.34 & 2.58 $\pm$ 0.03 & 0.47 & 0.248 & 0.593 & 9.12 & 0.56\\[0.0cm] 
KGP\_BET(B)\_$D^*10$nm & 0.18 & 1.91 $\pm$ 0.02 & 0.26 & 0.247 & 0.429 & 9.12 & 0.56\\[0.0cm] 
KGP\_BET(B)\_$D^*80$nm & 0.22 & 1.96 $\pm$ 0.02 & 0.30 & 0.247 & 0.465 & 9.12 & 0.56\\[0.0cm] 
\hline
\multicolumn{8}{l}{$^{\dag}$ Averaged over 121 local nanoindentation tests (distributions shown in the figures below)}\\[-0.1cm]
\multicolumn{8}{l}{$^{\dag\dag}$ Averaged over 6 tests for NaPG (2 BET tests for each of 3 MIP tests) and over 3 MIP tests for KGP}
\end{tabular}%
}
\label{tab_indent}
\end{table}

For our KGP samples, the porosimetry analysis in the previous section has indicated that any $D^*$ between 10 and 80 nm could be equally suitable. Therefore, we have run the micromechanical model for both $D^* =10$ nm and $D^* =80$ nm. The two BET tests, BET(A) and BET(B), yielded significant differences, thus we computed $\varphi_2$ from both tests separately. The results of all these analyses are collected in \tabname\ref{tab_indent}. The three MIP tests performed on KGP samples yielded instead similar results (see \figname\ref{fig_pores}); since MIP contributes to $\varphi_2$ only through the skeletal density $\rho_s$, \tabname\ref{tab_indent} shows $\rho_s$ averaged over the three MIP tests.

The results for KGP in \tabname\ref{tab_indent} show that the micromechanical model systematically estimates greater average total porosities $\phi$ than the porosimetry analysis in \figname\ref{fig_poroskele}(D).Two considerations allow us to identify the most representative analysis out of the four for KGP reported in \tabname\ref{tab_indent}. First, BET(B) is probably underestimating $v_{BET}(D^*)$ and thus also $\rho_s$, $\varphi_2$ and eventually $\phi$; this is because BET(B) did not sample pores smaller than 3 nm, whereas BET(A) reached down to 1 nm, as already discussed in relation to \figname\ref{fig_pores}(F). Second, both the analyses from the BET(A) dataset indicated a similar total volume of pores smaller and greater than $D^*$; these volumes are proportional to $\varphi_1$ for the capillary pores and to $\phi-\varphi_1$ for the smaller mesopores. For $D^* = 10$ nm, \tabname\ref{tab_indent} reports $\varphi_1 = 0.247$ and $\phi-\varphi_1 = 0.278$; for $D^* = 80$ nm, $\varphi_1 = 0.248$ and $\phi-\varphi_1 = 0.345$. From the porosimetry results in \figname\ref{fig_pores}(F), a similar volume of pores smaller and greater than $D^*$ is acceptable for $D^* = 10$ nm but not for $D^* = 80$ nm, because pores smaller than 80 nm should be largely predominant. All this leads us to consider the KGP dataset from BET(A) and with $D^* = 10$ nm as most representative, and thus focus on it for deeper analysis. 

\begin{figure}
    \centering
    \includegraphics[width=1\linewidth]{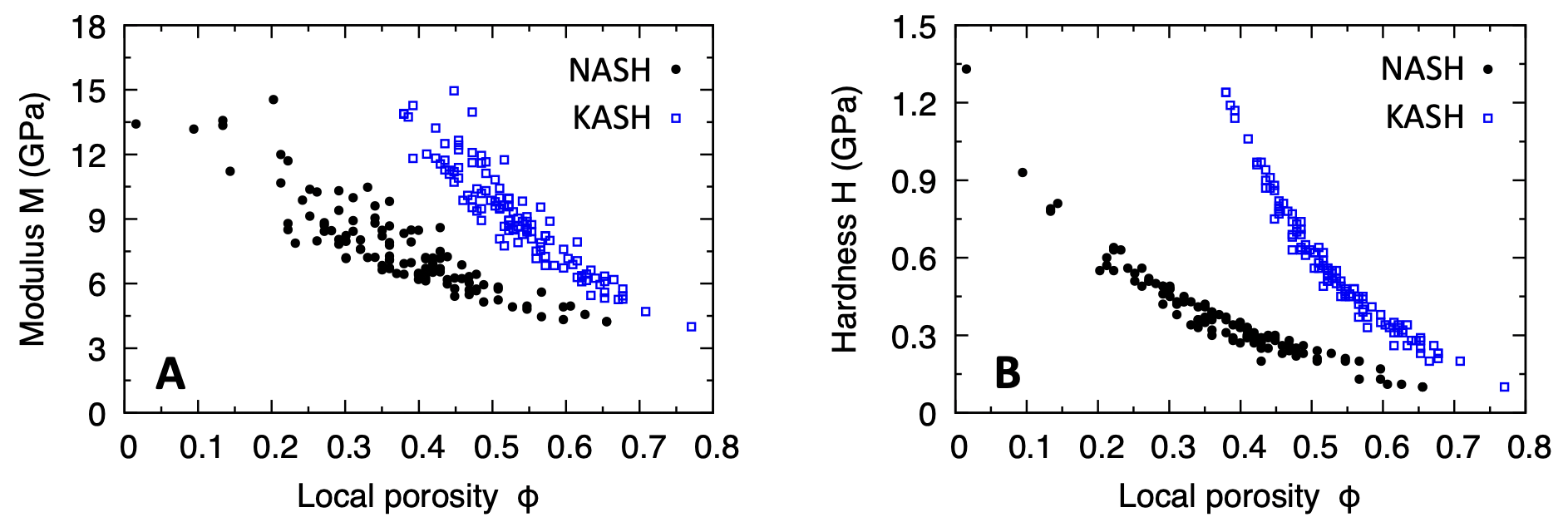}
   \caption{Experimental results from instrumented nanoindentation, (A) modulus and (B) hardness, associated to local porosities $\phi$ thorugh the micromechanical model in \secname\ref{secIndent}. The data for KASH are those based on the BET(A) dataset with $D^*=10$ nm.
    }
    \label{fig_nanoindent}
\end{figure}

For the selected KGP dataset from BET(A) and $D^* = 10$ nm, the skeletal density from porosimetry is \ca 2.29 g/cm$^3$ and the estimated total porosity from the micromechanical model is $\phi = 0.525$, which is \ca 10\% greater than the upper bound from porosimetry in \figname\ref{fig_poroskele}(D). \cite{kriven2006microstructure} also measured greater porosity in KGP than in NaGP, although by a smaller margin: \tabname\ref{tab_indent} reports $\phi=0.525$ and 0.373 respectively, meaning 40\% more porosity in KGP; \cite{kriven2006microstructure} measured 0.29 mL/g \vs 0.23 mL/g , so 26\% more in KGP. Despite the greater average porosity, the indentation tests measured better average mechanical properties for the KASH gel in the KGP: modulus $M = 9.12$ GPa (well in the 8-10 GPa most probable range previously reported in \cite{akono2019influence}) \vs 7.5 GPa for the NASH gel in the NaGP (22\% greater), and hardness $H = 0.56$ GPa \vs 0.37 GPa for NASH (51\% greater). \figname\ref{fig_nanoindent} shows the results of all 121 indentation tests for both our NASH and KASH gels, highlighting the wide spread of values for mechanical properties and local porosities, as well as the negative correlation between them arising from the micromechanical model. For any given $\phi$ value in \figname\ref{fig_nanoindent}, KASH displays higher $M$ and $H$ than NASH, indicating that the solid skeleton of KASH should be inherently stiffer and stronger. This interpretation is consistent with the skeletal density $\rho_s$ of KASH being 15\% greater than that of NASH (2.29 g/cm$^3$ \vs 2.02 g/cm$^3$ in \tabname\ref{tab_indent}). The results from atomistic simulations in the next section corroborate this point.

\subsection{Atomistic simulations}\label{secResAtomS}

\begin{figure}
    \centering
    \includegraphics[width=0.8\linewidth]{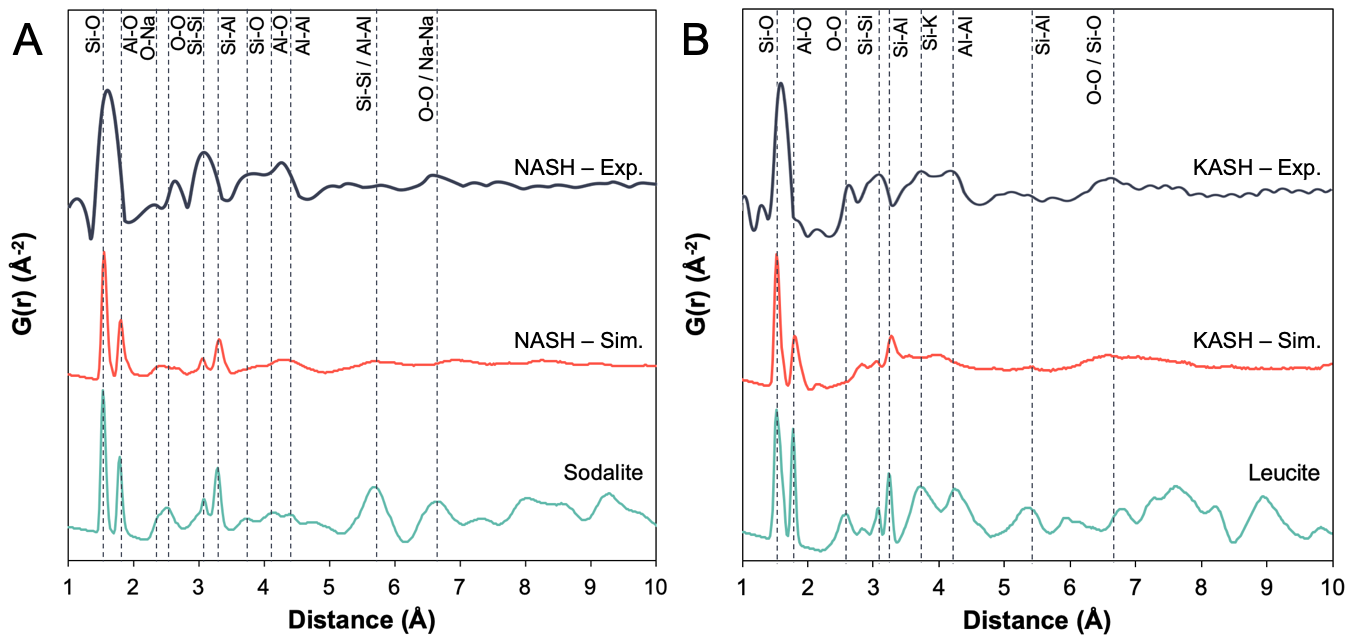}
    \caption{X-ray Pair Distribution Function (PDF) of (A) experimental and simulated NASH and sodalite, and (B) experimental and simulated KASH and leucite.}
    \label{fig:PDF}
\end{figure}

The defective structures of NASH and KASH are analysed by computing their X-ray Pair Distribution Function (PDF) and comparing them with the X-ray PDF of the starting crystalline structures, sodalite and leucite, respectively, and those of experimental samples of NaGP \citep{white2013situ} and KGP \citep{bell2008x}: see \figname\ref{fig:PDF}.
The diffraction patterns of sodalite and leucite feature well-defined peaks, typical of crystalline materials with ordered and periodic atomic arrangements. The positions of these peaks correspond to the interatomic distances within the crystalline lattice, whilst their intensities are related to the periodicity and the quantity of atoms in the structure. Our models of geopolymer structures also feature well-defined peaks at short distances (r < 4 \AA) at positions corresponding to those of the peaks in the parent crystals (sodalite and leucite). This indicates the persistence of some short-range order in the geopolymer models, despite the defects that have been introduced. However, at larger distances, the diffraction pattern in the model structures becomes more diffuse, without peaks, which is typical of amorphous materials. This combination of short-range order and long-range disorder is indeed confirmed by the experimental pattern in \figname\ref{fig:PDF}. Nevertheless, there are some differences between the simulated and experimental spectra; one particularly marked discrepancy concerns the peaks at around 1.7 \AA. The experimental PDF displays a single broad peak at 1.7 \AA, whereas the simulated geopolymers and their parent crystal structures feature two narrow peaks. This discrepancy, which already emerged in previous simulations of the NASH gel \citep{lolli2018atomistic}, may be attributed to experimental limitations in deconvoluting the contributions of Si and Al to that peak, or vice versa to the simulations overconstraining the atomic positions of Si and Al around local equilibrium points giving rise to two sharp peaks. 

\tabname\ref{tab:SkeletalDensities} shows the computed skeletal densities $\rho_s$ for NASH and KASH structures with Si:Al ratios of 1.4 and 2.0, also compared to all-siliceous model structures with Si:Al = $\infty$. The all-siliceous structures have the lowest $\rho_s$ because the absence of Al in them implies that there are no Na$^+$ or K$^+$ ions, nor are there water molecules from their hydration shells; consequently, their porous frameworks are empty. Nevertheless, the models with Si:Al = $\infty$ are instructive as they show that only a small difference in $\rho_s$ arises from the different structural arrangements in our NASH and KASH gel models: 2.13 vs. 2.09 g/cm$^3$, so just 2\%. In the model structures with Si:Al = 1.4 and 2.0, the presence of Na$^+$ and K$^+$ ions and of water molecules in their solvation shells, lead to higher $\rho_s$ values and a greater difference between NASH and KASH, with the latter being denser by 5-7\%. This result might seem counterintuitive since our model NASH structures contain more water per ion than those of KASH. The explanation is that both our NASH and KASH models tend to shrink when equilibrated to zero average stress after having substituted some Si with Al and after having added ions and water; water and ions actually reduce this tendency to shrink, thus the model of NASH (which has more water) shrinks less than KASH. Another factor contributing to the greater mass of KASH is also the greater atomic mass of K compared to Na ions.
\begin{table}
\caption{Computed skeletal densities $\rho_s$ (in g/cm$^3$) of NASH and KASH model structures. Si:Al = $\infty$ refers to the all-siliceous versions of the model structures. 
}

\centering
{%
\begin{tabular}{c c c c}
\hline
Si:Al & 1.4 & 2.0 & $\infty$\\
\hline
NASH & 2.38 & 2.32 & 2.09\\
KASH & 2.51 & 2.49 & 2.13\\
\hline
\end{tabular}%
}
\label{tab:SkeletalDensities}
\end{table}

The greater $\rho_s$ of KASH predicted by our atomic models is in line with our experimental findings from porosimetry in \secname\ref{secResPoro}, which indicated $\rho_s = 1.97-2.07$ g/cm$^3$ for NaGP and $\rho_s = 2.0-2.6$ g/cm$^3$ for KGP, \ie KASH having a \ca 15\% greater $\rho_s$ than NASH. In absolute values, our simulated $\rho_s$ for KASH are within the experimental range, close to its upper bound. By contrast, the simulations for NASH predict significantly higher values than our experiments. However, other experimental values of $\rho_s$ for NaGP available in the literature extend the range obtained in this article, with \cite{duxson2007physical} indicating $\rho_s$ as low as 1.7-1.8 g/cm$^3$, and \cite{nvemevcek2011nanoindentation} indicating $\rho_s = 2.372$ g/cm$^3$; this last value is indeed close to the simulation predictions in \tabname\ref{tab:SkeletalDensities}. Overall, the comparison between simulated and experimental skeletal densities is satisfactory, especially when considering that the simulation results are straight predictions, with no fitting from experimental data.

\begin{figure}
    \centering
    \includegraphics[width=0.5\linewidth]{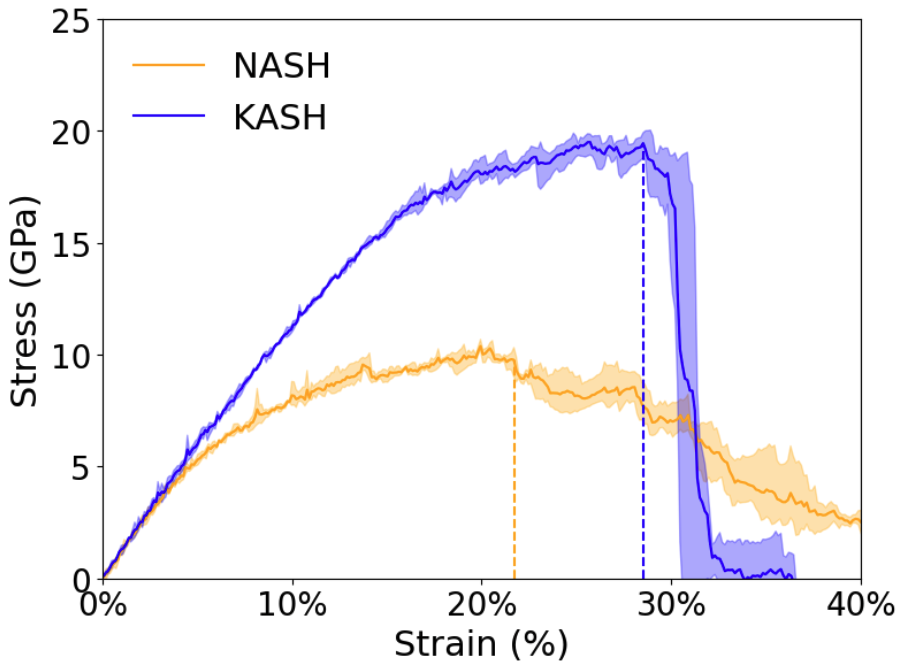}
    \caption{Stress-strain curves of atomic models of NASH and KASH gels with Si:Al = 2.0, under uniaxial strain. The dashed lines indicate the tensile strain at rupture. The shaded areas encompass the limits of three repeated simulations. 
    }
    \label{fig_TSR}
\end{figure}

\figname\ref{fig_TSR} shows the simulated stress-strain curves for our atomic structures with Si:Al = 2 under uniaxial strain. The initial slope of the curves quantifies the indentation modulus $M$, whose values are shown in \tabname\ref{tab:MechanicalProperties}. Since all the simulated structures have similar Poisson's ratios $\nu$, the elastic moduli, $E$, are all proportional to the corresponding $M$.
%
%
$E$ and $M$ are greater for KASH than for NASH (by 17\%) and  greater for low Si:Al ratios (by 11\%); this qualitatively matches the trend in skeletal density from \tabname\ref{tab:SkeletalDensities}. 
\begin{table}
\caption{Elastic constants \textcolor{black}{and their corresponding standard errors} computed from the atomic models of NASH and KASH (averages over mechanical tests in the three Cartesian directions).}
\centering
\begin{tabular}{c c c c c c c}
\hline
& \multicolumn{2}{c}{E (GPa)} & \multicolumn{2}{c}{$\nu$} & \multicolumn{2}{c}{M (GPa)}\\
\hline
Si:Al & 1.4 & 2.0 & 1.4 & 2.0 & 1.4 & 2.0\\
\hline
NASH & 61.58 {\color{black}$\pm$  2.00} & 59.42 {\color{black}$\pm$ 1.60} & 0.33 {\color{black}$\pm$ 0.0025} & 0.31 {\color{black}$\pm$ 0.0042}& 92.23 {\color{black}$\pm$ 2.95} & 82.81 {\color{black}$\pm$ 2.20}\\
KASH & 71.61 {\color{black}$\pm$ 0.96} & 65.97 {\color{black}$\pm$ 1.11} & 0.34 {\color{black}$\pm$ 0.0027} & 0.33  {\color{black}$\pm$ 0.0046} & 108.31 {\color{black}$\pm$ 1.40} & 96.67 {\color{black}$\pm$ 1.53}\\    
\hline
\end{tabular}%
\label{tab:MechanicalProperties}
\end{table}
\figname\ref{fig_TSR} also highlights a very nonlinear behaviour of our atomic models at high strain levels, eventually quantifying the stress and strain at failure, $\sigma_u$ and $\varepsilon_u$, both of which contribute to the hardness of the materials at larger length scales. \figname\ref{fig_TSR} shows $\varepsilon_u = 0.22$ and 0.28 respectively for NASH and KASH gels with Si:Al = 2.0. These values, along with the elastic moduli for the gels with the same Si:Al, are used to parametrise the stiffness and cutoff of the interaction potentials between nanoparticles at the mesoscale. \figname\ref{fig_TSR} also shows that the failure stress of KASH, $\sigma_u = 19.5$ GPa, is approximately twice that of NASH, $\sigma_u = 10$ GPa. The greater stiffness ($M$ and $E$) and failure stress of KASH predicted by our simulations confirm our previous interpretation of the experimental results from nanoindentation, \ie that KASH has an intrinsically stiffer and solid solid skeleton than NASH. Also the fact that the difference in simulated strength (twice as high for KASH) is greater than that in simulated $M$ (17\% higher for KASH) agrees qualitatively with the greater difference in hardness than in modulus shown in \figname\ref{fig_nanoindent}.


\subsection{Mesoscale simulations}\label{secResMesoS}

\begin{figure}
    \centering
    \includegraphics[width=0.5\linewidth]{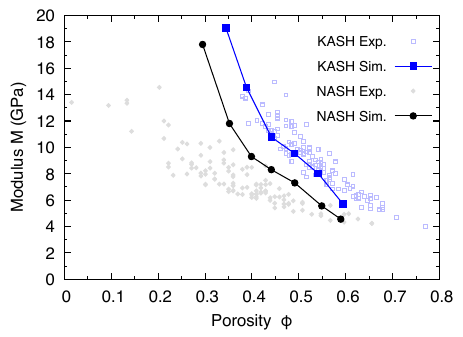}
    \caption{Nanoindentation moduli from mesoscale particle-based simulations compared to the experimental data from \figname\ref{fig_nanoindent}(A). The porosities are those of the model gel structures after initial equilibration to zero stress.}
    \label{fig_Computed_M}
\end{figure}

\figname\ref{fig_Computed_M} compares the experimental data on indentation moduli $M$ from \figname\ref{fig_nanoindent}(A) with the results from our particle-based mesoscale simulations of NASH and KASH mechanics. In the 0.4 -- 0.6 porosity range, there is good agreement between experiments and simulations, except that the simulations tend to overpredict $M$ for NASH. For $\phi < 0.4$, the simulations overestimate $M$ certainly for NASH and probably for KASH too (for KASH we do not have experimental data for $\phi < 0.38$). Nevertheless, the agreement between simulations and experiments in \figname\ref{fig_Computed_M} is quite remarkable considering that the simulation results are straight predictions, since the only experimental input for the simulations was the particle size distribution, whereas the mechanical interactions were fully parametrised from the results of our atomistic simulations. Hardness values were not computed for the model mesostructures because the adopted harmonic interaction potentials (\viz perfectly brittle linear springs) cannot capture the large nonlinearity of the stress-strain curves from the atomistic simulations, in \figname\ref{fig_TSR}. Had we employed more sophisticated interactions, the mesoscale simulations have probably captured the significantly higher hardness of KASH compared to NASH, since the atomistic simulations in \figname\ref{fig_TSR} predicted indeed much higher strength and failure strain for KASH.

\begin{figure}
    \centering
    \includegraphics[width=1\linewidth]{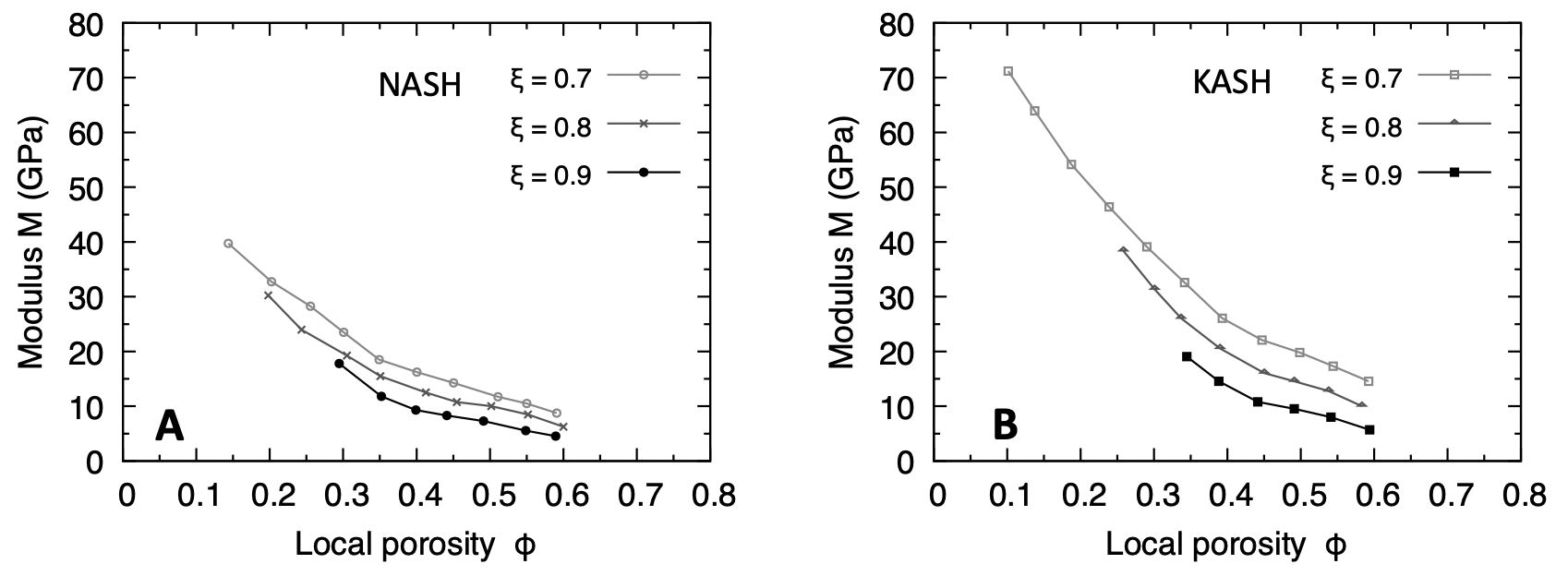}
    \caption{Indentation modulus \vs porosity curves from simulations on our model mesostructures of (A) NASH and (B) KASH, for different values of the parameter $\xi$ that controls the allowed overlapping between particles, and therefore the intensity of the eigenstresses generated when building the model structures.}
    \label{fig_cred}
\end{figure}

The overestimated $M$ values from our simulations at low porosities are the result of a sharp change in gradient of the simulated $M(\phi)$ curves when $\phi$ decreases below 0.4 -- 0.45. This range of porosity is close to the geometric Random Loose Packing limit for monodisperse spheres ($\phi = 0.46$ and $\eta = 1 - \phi = 0.54$), which suggests that the change in $M(\phi)$ gradient might correspond to a marked increase in the number of interacting neighbours per particle (\viz coordination) and of local eigenstress intensity. In our simulations, coordination and local eigenstresses are interrelated, since a porosity below the geometric packing limit can only be reached by slightly pushing the exiting particles out of local mechanical equilibrium (which creates eigenstress) and thus making room for additional particles in the interstices (which increases the coordination).
A thorough analysis of the role of coordination and eigenstresses is beyond the scope of this article. However, to substantiate the proposition that local eigenstresses may significantly impact the simulated elastic properties, \figname\ref{fig_cred} shows families of $M(\phi)$ curves for our mesoscale models of NASH and KASH obtained by adopting different $\xi$ values during the packing process to create the model structures: see \secname\ref{secSimCGMet}. $1-\xi$ is the fraction of radial overlapping that is allowed between a newly inserted particle and already existing ones. $\xi = 1$ means that no overlapping is allowed and, therefore, new particles cannot push existing ones away to make room for themselves during packing. By contrast, smaller values of $\xi$ imply that new particles can be added even if they partly overlap with existing ones; when this happens, the structure relaxes to zero average axial stresses still retains residual eigenstresses whose intensity is larger the smaller $\xi$ is \citep{masoero2020nanoparticle}. \figname\ref{fig_cred} shows that allowing for overlapping during packing (\via a small $\xi$) makes it possible to reach higher packing densities, \viz smaller porosities; compare the leftmost point of the various curves in the figure, which is the maximum packing we managed to achieve in the various cases. \figname\ref{fig_cred} also shows that, for any given porosity $\phi$, the curves obtained using a smaller $\xi$ (\viz with greater eigenstresses) display indeed greater moduli $M$.
The dependence of elastic moduli on eigenstresses is a known phenomenon \citep{zhang2022prestressed} that has practical consequences, since eigenstresses are the result of the gel formation kinetics and therefore they depend on the synthesis process. This means that synthesis variables, such as curing time and temperature, may significantly impact the elastic properties of the resulting gel phases and, ultimately, of the geopolymer paste.

\section{Conclusion}

The article has compared the microstructure and micromechanics of NASH and KASH gels in sodium and potassium based geopolymers from metakaolin, NaGP and KGP. The synergy between experimental findings (from porosimetry, microscopy, and nanoindentaiton) and predictive simulations (atomistic and coarse-grained) has confirmed some previous findings from the literature and added new insights.

TEM microscopy has confirmed the nanogranular structure of both NASH and KASH gels, with NASH featuring coarser particles of \ca 30 nm in size as opposed to \ca 5 nm in KASH. To characterise the multiscale porosity of the gels, we have combined mercury intrusion porosimetry (MIP) and nitrogen gas sorption (BET), showing that the porosity in KGP is dominated by pores with characteristic size of \ca 5 nm, whereas most pores in NaGP are in the 20 -- 80 nm range. We also argued how an ink-bottle effect may cause MIP to overestimate the volume of capillary pores in NaGP.

Through porosimetry we also measured skeletal density $\rho_s$ and total porosity $\phi$. To manage some mismatch between MIP and BET in the 10 -- 80 nm range, we computed $\rho_s$ and $\phi$ for various scenarios obtained considering various pore diameter thresholds $D^*$ below which we trusted BET better than MIP, and vice versa above $D^*$. The analysis indicated a higher skeletal density $\rho_s$ for KASH ($2.3 \pm 0.3$ g/cm$^3$) than for NASH ($2.02 \pm 0.05$ g/cm$^3$), a total porosity $\phi = 0.38 - 0.39$ for NASH and an inconclusively wide porosity range (0.32 -- 0.48) for KASH. Experimental results from nanoindentation, interpreted through a micromechanical model, corroborated the upper bound value of porosity for KASH, indicating $\phi$ as high as 0.525.

The indentation experiments showed that KASH has higher average modulus $M$ and hardness $H$ than NASH: $M=9.12$ GPa and $H=0.56$ GPa for KASH \vs 7.5 GPa and 0.37 GPa for NASH. This result, along with the higher $\rho_s$ and $\phi$ of KASH, indicated that the solid skeleton of KASH should be inherently stronger and stiffer than in NASH. This hypothesis was corroborated by molecular mechanics simulations based on a recent model of NASH \citep{lolli2018atomistic} and on a new model of KASH (defective leucite), which predicted indeed higher skeletal density, modulus, and tensile strength for KASH: $\rho_s = 2.49$ g/cm$^3$, $M = 96.7$ GPa, $\sigma_u = 19.5$ GPa, whereas for NASH, $\rho_s = 2.32$ g/cm$^3$, $M = 82.8$ GPa, and $\sigma_u = 10$ GPa.

Coarse grained, particle-based, model structures of NASH and KASH were thus developed, featuring a range of porosities $\phi$ and harmonic interaction potentials between particles that were fully parametrised from the atomistic simulations. The resulting mesoscale simulations predicted relationships between $M$ and $\phi$ that matched the corresponding experimental data from nanoindentation for KASH and NASH in the 0.4 -- 0.6 porosity range. In particular, the simulations confirmed that, for any given porosity $\phi$, the KASH gel is indeed expected to have better mechanical properties than NASH, owing to a stiffer and stronger interaction potential. For $\phi < 0.4$ the mesoscale simulations overpredicted the moduli for both NASH and KASH, which may be due to artificial eigenstresses in the model structures. Further simulation efforts could target a better and quantitative understanding of residual eigenstresses in geopolymer paste, as a way to explain, at least in part, the impact of the synthesis protocol on the micro-mechanical properties of these binders.

\bigskip
\textbf{Acknowledgments}

E.D-R. acknowledges the support of the Basque Government through the IT1639-22 project and the postdoctoral fellowship from “Programa Posdoctoral de Perfeccionamiento de Personal Investigador Doctor”. We gratefully acknowledge the technical and human support provided by SGIker (UPV/EHU - ERDF, EU), for the allocation of computational resources provided by the Scientific Computing Service.

A.T.A. acknowledges the support of the National Science Foundation under grant number DMR 1928702. This work made use of the EPIC facility of Northwestern University’s NUANCE Center, which has received support from the SHyNE Resource (NSF ECCS-2025633), the IIN, and Northwestern's MRSEC program (NSF DMR-2308691).

E.K. has participated in this project through the contribution of the Royal Society and the African Academy of Science, funding FLAIR. Grant FLR\textbackslash R1\textbackslash 201402. 


\bibliographystyle{elsarticle-harv}
\bibliography{./bibliocement}


\end{document}